\documentclass[twocolumn,tighten,times]{aastex631}

\usepackage{amsmath}

\accepted{AJ}

\shorttitle{Exploring the Magnetic Field Geometry in NGC 891 with SOFIA/HAWC+}
\shortauthors{Kim, Jones \& Dowell}
\graphicspath{{./}}
\begin{document}

\title{Exploring the Magnetic Field Geometry in NGC 891 with SOFIA/HAWC+}

\author[0000-0001-8380-9988]{Jin-Ah Kim}
\affil{Minnesota Institute for Astrophysics, University of Minnesota, Minneapolis, MN 55455, USA}

\author[0000-0002-8716-6980]{Terry Jay Jones}
\affil{Minnesota Institute for Astrophysics, University of Minnesota, Minneapolis, MN 55455, USA}

\author{C. Darren Dowell}
\affil{NASA Jet Propulsion Laboratory, California Institute of Technology, 4800 Oak Grove Drive, Pasadena, CA 91109, USA}



\begin{abstract}

SOFIA/HAWC+ $154~\micron$ Far-Infrared polarimetry observations of the well-studied edge-on galaxy NGC 891 are analyzed and compared to simple disk models with ordered (planar) and turbulent magnetic fields. The overall low magnitude and the narrow dispersion of fractional polarization observed in the disk require significant turbulence and a large number of turbulent decorrelation cells along the line-of-sight through the plane. Higher surface brightness regions along the major axis to either side of the nucleus show a further reduction in polarization and are consistent with a view tangent to a spiral feature in our disk models. The nucleus also has a similar low polarization, and this is inconsistent with our model spiral galaxy where the ordered magnetic field component would be nearly perpendicular to the line-of-sight through the nucleus on an edge-on view. A model with a barred spiral morphology with a magnetic field geometry derived from radio synchrotron observations of face-on barred spirals fits the data much better. There is clear evidence for a vertical field extending into the halo from one location in the disk coincident with a polarization null point seen in near-infrared polarimetry, probably due to a blowout caused by star formation. Although our observations were capable of detecting a vertical magnetic field geometry elsewhere in the halo, no clear signature was found. A reduced polarization due to a mix of planar and vertical fields in the dusty regions of the halo best explains our observations, but unusually significant turbulence cannot be ruled out.

\end{abstract}



\section{Introduction} \label{sec:intro}

Magnetic fields are ubiquitous in spiral galaxies and an important component to the dynamical picture of the interstellar medium. The magnetic fields are weak in the early universe but get amplified inside spiral galaxies \citep[see][for review]{2015A&ARv..24....4B}. Differential rotation in galaxies intensifies the strength of the magnetic fields by the large-scale dynamo and ordering turbulent magnetic fields \citep{1988Natur.336..341R}. The turbulence of magnetic fields is likely caused by processes such as cloud collapse and supernova explosions \citep[e.g.,][]{2012MNRAS.422.2152B, 2013A&A...560A..87S}. However, the question of the generation and evolution of the global geometry of magnetic fields in galaxies is still uncertain. 

The magnetic field geometry in galaxies and the Milky Way has been studied by observing polarization due to dichroic extinction of starlight at optical and near-infrared (NIR) wavelengths \citep[e.g.,][]{1970MmRAS..74..139M, 1997AJ....114.1393J, 2014ApJ...786...41M}, dichroic emission at far-infrared (FIR)/sub-mm wavelengths \citep[e.g.,][]{2021MNRAS.505..684P, 2022ApJ...936...92L}, and polarized synchrotron radiation at radio wavelengths \citep[see][for review]{2015A&ARv..24....4B}. Polarized light (in transmission) observed at optical/NIR wavelengths in large beams on external galaxies can easily be contaminated by scattered light entering the beam \citep[e.g.,][]{1997AJ....114.1405W}. Synchrotron radiation from radio emission in galaxies is easily influenced by Faraday rotation \citep{1963Natur.197.1162G}, especially in a galactic disk in edge-on galaxies. To infer magnetic field orientation on the plane of the sky, the effect of Faraday rotation can be corrected based on multi-frequency radio polarization observations. However, the correction can be challenging, especially for an edge-on disk with a large column depth consisting of multiple regions in a beam. Both scattering at short (optical/NIR) wavelengths and Faraday rotation at long (radio) wavelengths affect the position angle of the net polarization, and hence obscure the underlying magnetic field geometry. At FIR wavelengths, neither scattering nor Faraday effects are significant factors. Note that interstellar polarization does not provide information on the +/- sign of the magnetic field, just the orientation in the plane of the sky. Faraday rotation and Zeeman splitting can be used to examine the magnetic field sign along a line-of-sight (LOS) \citep{2013pss5.book..641B}.

Magnetic fields in many spiral galaxies inferred from synchrotron radiation appear as spiral patterns on a galactic plane, and the field lines in the halo generally extend outward from a galactic plane, often forming an X-shaped pattern in edge-on systems \citep{2015A&ARv..24....4B}. NIR and FIR polarimetry reveal that the magnetic fields near a galactic plane lie primarily in the plane \citep[e.g.,][]{1997AJ....114.1393J, 2020AJ....160..167J}. However, magnetic fields observed in the halo at these wavelengths are found to be more complicated \citep[e.g.,][]{2000AJ....120.2920J, 2019ApJ...870L...9J, 2021MNRAS.505..684P}. The $\alpha$-$\Omega$ theory \citep{1988Natur.336..341R}, commonly used to explain the magnetic field geometry in galaxies, assumes a thin disk and does not consider the field in the halo. The typical magnetic field geometry and the existence of vertical fields in a halo are still a subject of debate. Observing edge-on galaxies is ideal for studying mechanisms for the formation and evolution of the magnetic field geometry in the halo since the much brighter disk emission is absent. Continuum Halos in Nearby Galaxies-an EVLA Survey \citep[CHANG-ES;][]{2012AJ....144...43I} is one recent project studying magnetic fields observed in edge-on galaxies with radio observations. However, observing polarization at FIR wavelength has the advantage of disregarding the effects of scattering and Faraday rotation which may significantly affect polarized light observed in edge-on galaxies. Vertical fields in a halo have been seen at FIR wavelength observations only in M82 \citep{2019ApJ...870L...9J} and NGC 2146 \citep{2022ApJ...936...92L}, which are starburst galaxies, and Cen A, which is a late merger system with an active AGN \citep{2021NatAs...5..604L}. Vertical fields are observed prominently at shorter wavelengths, which may be associated with warmer dust, than at longer wavelengths in M82 \citep{2021MNRAS.505..684P} and NGC 2146 \citep{2022ApJ...936...92L}.

FIR polarimetry has proven to be a useful tool for tracing the magnetic field geometry in the Milky Way and external galaxies \citep[e.g.,][]{1982MNRAS.200.1169C, 1988QJRAS..29..327H}. FIR emission is normally caused by thermal radiation from warm dust heated by the interstellar radiation field. The dust grains are elongated in shape and align with the local magnetic field \citep{2007JQSRT.106..225L}, resulting in polarized emission along the long axis of the grain and perpendicular to the orientation of the magnetic field. A net alignment is necessary as polarization cannot be observed if the axes of the dust grains are randomly distributed. Radiative Torque alignment (RATs) is the current leading model for grain alignment and is discussed in \cite{2015ARA&A..53..501A}. FIR polarimetry measures the magnetic field geometry projected on the plane of the sky, and the fractional polarization depends on the orderliness of the projected magnetic fields, the inclination of magnetic field geometry from the LOS, and the degree of dust alignment. In addition to the projected field geometry, we can estimate a field strength \citep[e.g.,][]{1951PhRv...81..890D, 1953ApJ...118..113C, 2011ApJ...733..109H} and obtain information on the turbulence in the field along a LOS \citep{1992ApJ...389..602J, 2011ApJ...733..109H}. Importantly, FIR polarimetry samples regions with warm dust, unlike radio synchrotron emission, which samples all regions with cosmic ray electrons.

NGC 891 is a nearby edge-on spiral galaxy at a distance of 8.36 Mpc \citep{2001ApJ...546..681T} and a star formation rate of 3.8 $M_{\odot}~\text{yr}^{-1}$ \citep{2004A&A...414...45P}. The polarization in NGC 891 has been studied using optical, NIR, FIR, and radio observations \citep[e.g.,][]{1996A&A...308..713F, 2014ApJ...786...41M, 2020AJ....160..167J, 2020A&A...639A.112K}. \citet{1996A&A...308..713F} observed many polarization lines oriented perpendicular to the galactic disk. As discussed in \cite{ 2020AJ....160..167J}, the optical polarization is probably dominated by scattering and does not accurately represent magnetic field geometry. The fractional polarization in NGC 891 at NIR wavelengths was found to be significantly lower than expected compared to the Milky Way and other nearby spirals such as NGC 4565 \citep{1997AJ....114.1393J}. \citet{1997AJ....114.1393J} suggested either there were regions of star formation producing large-scale blowouts from the plane that created a vertical component to the magnetic field geometry or that there was turbulence in the field on much smaller scales than expected. Blowouts would imply that the halo region near the disk should have a vertical field geometry. Observations at NIR wavelengths \citep{1997AJ....114.1393J, 2014ApJ...786...41M} reveal that the average position angle of the polarization vectors in the disk plane is 12$\degr$ westward (clockwise on the sky) from the major axis of the disk. \citet{2014ApJ...786...41M} find evidence in the northeast part of the disk of a polarization null point, where the NIR fractional polarization is close to zero. This was tentatively associated with a LOS down an embedded spiral feature.

Dichroic extinction within the galactic plane is mingled with polarization due to scattering, and contamination by scattering at the polarization null points cannot be ruled out  \citep{1997ApJ...477L..25W, 2018ApJ...862...87S}. Polarization angles observed from synchrotron radiation vary from the southern to the northern disk \citep{2014ApJ...786...41M}, and show an X-shaped pattern in the halo \citep{2009RMxAC..36...25K}. Recent radio observations with better resolution find some vertical fields located in a few patchy regions in the halo \citep{2020A&A...639A.112K}. Radio synchrotron emission is strongly affected by Faraday depolarization in the edge-on disk. 

NGC891 was first observed in polarization at FIR wavelengths with SOFIA/HAWC+ by \citet{2020AJ....160..167J}. They found that the magnetic field geometry lies close to the plane, with no indication of the position angle offset seen in the NIR, and were unable to find clear evidence for vertical fields off the disk. In this study, we use additional observations of NGC891 with SOFIA/HAWC+ at $154\micron$ to improve the polarimetry off the galactic plane and extend the coverage to the North where \citet{2014ApJ...786...41M} find a polarization null point. In section \ref{sec:obs_result}, we will discuss the polarimetry results from the observations. The results in the regions off the galactic plane are in section \ref{subsec:offplane} and compared with radio observations. Section \ref{result:model_result} will concentrate on the polarimetry close to the galactic plane and comparison with our computational models. Section \ref{sec:discussion} will discuss our results and plausible mechanisms affecting magnetic fields in NGC891.

\section{Observation Data} \label{sec:data}

NGC891 was observed at 154$\micron$ with the 2.7m Stratospheric Observatory For Infrared Astronomy (SOFIA) telescope \citep{2018JAI.....740011T} using the High-resolution Airborne Wideband Camera-plus \citep[HAWC+;][]{2018JAI.....740008H} in 2017 (AOR 70\_0509\_3), 2018 (AOR 70\_0609\_1), and 2021 (AOR 09\_0067\_1). The observations have been made in chop-nod polarimetric imaging mode. The chop amplitudes and angles are 150$\arcsec$ and 245$\degr$ in AOR 70\_0509\_3 and AOR 70\_0609\_1, and 200$\arcsec$ and 300$\degr$ in AOR 09\_0067\_1. The total on-source exposure time is 2.78 hr. The FWHM beam size is $13.6\arcsec$, and the detector pixel size is 6.9\arcsec. 

We start from level 3 data in the IRSA SOFIA Archive, which have been demodulated and corrected for instrumental polarization. We combine the level 3 data using HAWC+ data reduction pipeline v1.3.0beta3 developed by SOFIA/HAWC+ science team. World coordinate system has been corrected by comparing with VLA images used in section \ref{subsec:offplane} and matching the locations of the bright peaks. Flux is calibrated with the aim of matching with NGC891 HAWC+ data taken by 2018. When combining all HAWC+ observations, we use our own method of flux integration in overlapping areas, instead of the Gaussian smoothing built in the pipeline. The observed grid is projected onto the output grid, and the fluxes in overlapped pixels of all observations with each output pixel are integrated. The fluxes are summed with weighting by the overlapped area divided by the inverse square of intensity error, and the weighted value is normalized after the summation. The method is slower, but does not degrade the angular resolution and reduces the correlations between nearby pixels compared to Gaussian smoothing. This minimizes underestimating the final noise due to smoothing and correlations between pixels. The y-axis of the output grid is set along the galactic plane \citep[22.9\degr;][]{2014A&A...565A...4H}. We use image pixel sizes of $6.8\arcsec$, $13.6\arcsec$, $20.4\arcsec$, and $27.2\arcsec$ for the output grid in order to achieve a signal-to-noise ratio (S/N) that is useful for a given intensity level. Hereafter in this paper, the term `pixel size' refers to an output grid pixel not the intrinsic detector pixel size or FWHM beam size.

In the pipeline, a reduced $\chi_{r}^{2}$ test is performed to determine if there are sources of extra error \citep[see][for details]{2011ASPC..449...50N, 2011ApJ...732...97D}. The reduced $\chi_{r}^{2}$ should be 1 in the data set with no extra error beyond the nominal errors. The reduced $\chi_{r}^{2}$ test estimates the factor for increasing the nominal error so that the reduced $\chi_{r}^{2}$ from the inflated errors becomes 1. The HAWC+ pipeline does find greater errors in the observations than expected. The cause of extra errors is not well understood, is seen on short timescales in the observations, and is probably systematic in origin.

The SOFIA/HAWC+ pipeline user manual mentions that the correction for instrumental polarization should be good to within 0.6\% for Q/I and U/I. In the high intensity regions in the disk of NGC 891, where the fractional polarization is very low, our observations in Figure \ref{fig:polmap} show very coherent polarization lines aligned with the disk for most locations. This indicates any unknown instrumental polarization would have to be effectively exactly aligned or exactly perpendicular to the disk to go undetected. Well outside the galaxy on blank sky, our observations show Stokes Q and U intensities are randomly distributed about zero, indicating there are no zero-point offsets in the Stokes Q and U data.

From the final maps of Stokes I, Q, and U, we compute the position angle and fractional linear polarization. The position angle is calculated from atan2(U, Q)/2, then rotated $90\degr$ to infer the orientation of the magnetic fields. In this paper, all position angles represent the orientation of the inferred magnetic fields, not observed polarization angles. FIR polarimetry produces `polarization lines' which delineate the orientation of magnetic field lines, but not a +/- direction along that orientation. Fractional polarization is $\sqrt{\textrm{Q}^2 + \textrm{U}^2}/\textrm{I}$, which creates a positive bias, since pure noise in Stokes Q and U will always produce a positive fractional polarization. In certain cases, the fractional polarization is debiased using the equation in the appendix in \citet{1974ApJ...194..249W}, which is derived based on the most probable values in the Rice distribution. In this paper, the fractional polarization after being debiased is notated as `debiased p'. If not, the values are the `observed p', without correction. Since our models contain the observation errors, we will often use the `observed p' for comparison with the models. The position angle contains no bias, and we will use almost all of the observed Q and U intensities with their accompanying errors, not just those corresponding to pixels with an S/N cut in `debiased p'.

\section{Observation Result} \label{sec:obs_result}

\begin{figure*}[t]
\centering
\includegraphics[width=.85\linewidth]{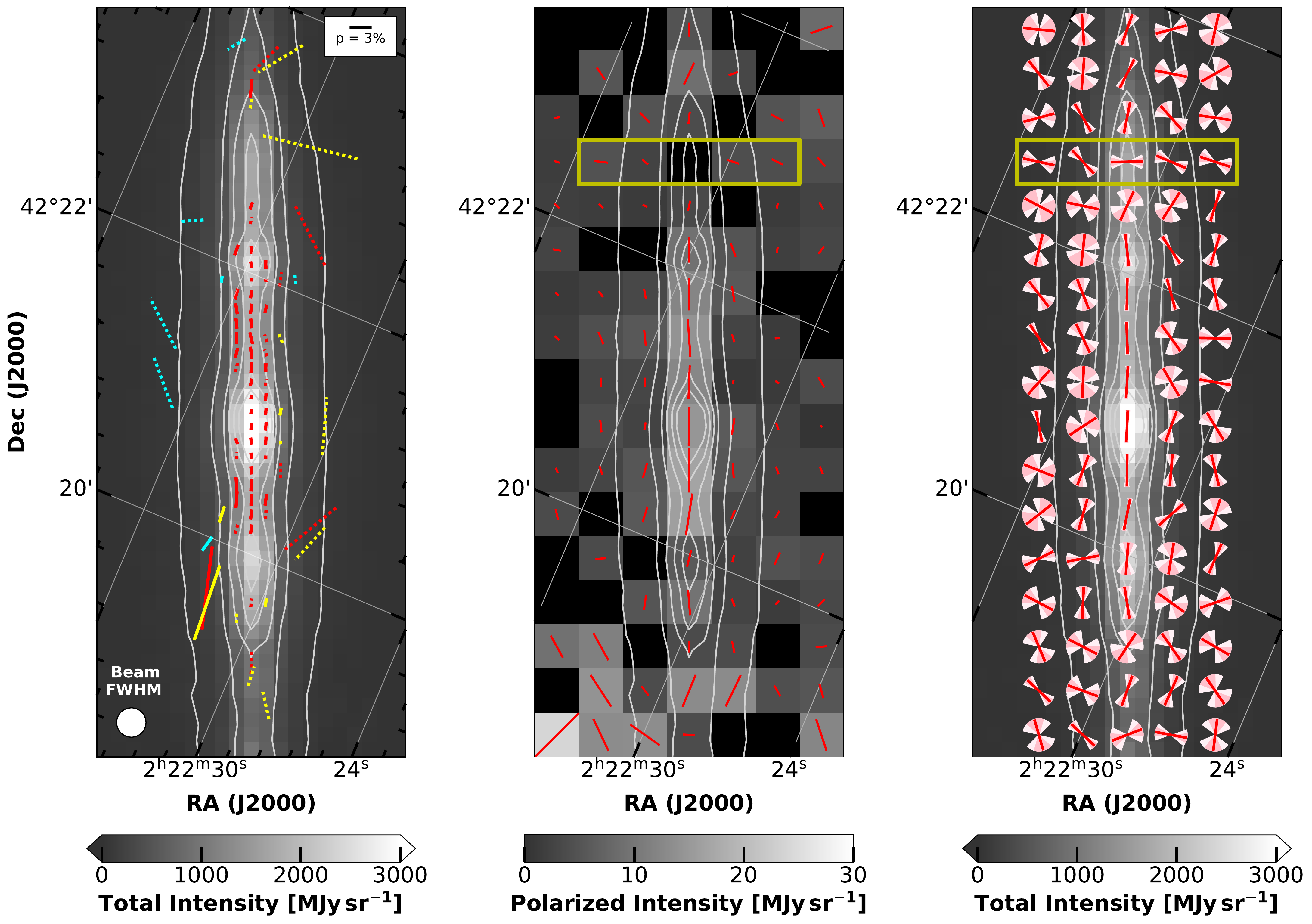} 
\caption{\textbf{Left}: Total intensity map, with a pixel size of 6.8\arcsec, overlayed with polarization lines. Different pixel sizes for the polarization lines are indicated by red (6.8\arcsec), yellow (13.6\arcsec), and cyan (27.2\arcsec) colors. The polarization lines have I/$\textrm{I}_{err}$ greater than 30. Solid lines indicate data with $p_{debiased}/p_{err} > 3$ and dotted line $3 > p_{debiased}/p_{err} > 2.5$. The line length  is proportional to the debiased fractional polarization. \textbf{Middle}: Polarization lines overlaying the debiased polarized intensity map with a pixel size of 20.4\arcsec. All polarization lines are shown regardless of S/N, but observed~$PI/PI_{err}$ has to be larger than 1 to be debiased. The line length is proportional to the debiased polarized intensity. \textbf{Right}: Position angle map computed within a pixel size of 20.4\arcsec. All polarization lines are represented regardless of S/N, and their lengths are plotted as a constant. Sectors of a circle in dark and bright pink show the extent encompassing 68.2\% and 90\% of samples based on Monte Carlo simulations. In all panels, contours are 100, 500, 1000, 1500, 2000, and 2500 MJy/sr in the total intensity map with a pixel size of 6.8\arcsec. The yellow boxed area delineates the vertical structure discussed in the text.} \label{fig:polmap}
\end{figure*}

Polarimetry for NGC891 observed with SOFIA/HAWC+ at $154\micron$ is shown in Figure \ref{fig:polmap}. All polarization lines in the figure represent the inferred magnetic field after being rotated 90\degr. The polarization lines in the left panel satisfy the criteria of having a S/N in total intensity (I/$\textrm{I}_{err} > 30$) and fractional polarization ($p_{debiased}/p_{err} > 2.5$). We vary the pixel size (different line colors) in locations with poorer S/N in the HAWC+ data. For a pixel where the S/N criteria are satisfied at the smaller pixel size, we use that pixel size. The right panel presents all vectors without any selection criteria within the area up to $51\arcsec$ off the galactic plane.

In the left panel, fractional polarization is low in the central region and increases with distance along the plane, but drops again at the brighter blobs to either side of the nucleus. The majority of the area away from the galactic plane has insufficient S/N to obtain reliable position angles. There are only a few vectors perpendicular to and well away from the galactic plane, suggesting some vertical fields may exist in the halo. We will discuss the possible presence of vertical fields in section \ref{dis:geo_off}.

The middle and right panel show all polarization lines without any cuts in S/N but using a pixel size of 20.4\arcsec (1.5 $\times$FWHM) to provide a uniform comparison. Using the pixel size larger than the beam size of 13.6$\arcsec$ is downsampling. However, we would like to distinguish intensities between the galactic plane and the halo using the beam size corresponding to $\sim$550 pc and improve S/N with a larger pixel size.

The middle panel shows polarization lines of which lengths are proportional to debiased polarized intensities. Because the debiasing method following the appendix in \citet{1974ApJ...194..249W} needs a S/N in polarized intensities larger than 1, the pixels having the S/N less than 1 are masked. Near the galactic plane, most polarization lines are strongly parallel to the disk all the way from the nucleus to the southern extent of the observations. The polarized intensities are the highest near the disk plane.

The right panel has polarization lines with a constant length, and the uncertainties in their position angles are layered. To estimate the uncertainty of the position angles, which can be quite large, we perform Monte Carlo simulations using Gaussian distributed errors in Stokes Q, and U with the observation errors taken as a standard deviation. The dark and bright pink sectors of a circle in the figure correspond to fractions of 68.2\% and 90\% out of 100,000 simulations. In the northern part of the galactic plane, there are two polarization lines that show a greater scatter in the observed position angle, even though the errors are small there.

\begin{figure}[t]
\centering
\includegraphics[width=\linewidth]{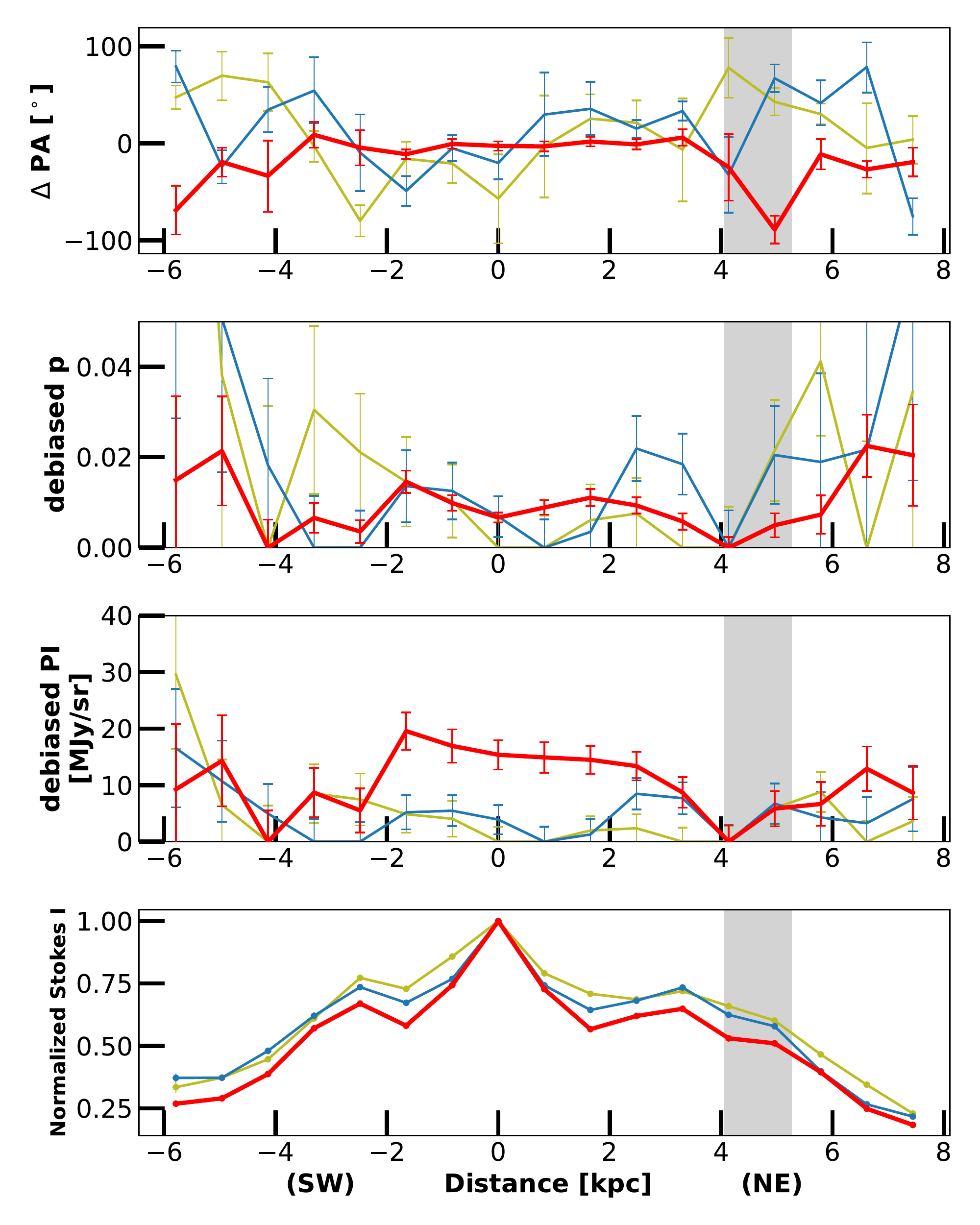}
\caption{Position angle, debiased fractional polarization, debiased polarized intensity, and total intensity as a function of the distance from the center along the major axis. Each data point corresponds to a polarization line shown in the right panel of Figure \ref{fig:polmap}, with a pixels size of 20.4$\arcsec$. Red is the midplane, and green and blue indicate the eastside and westside off 20.4$\arcsec$ about the galactic plane, respectively. Error bars in position angles are the distribution containing 68.2\% of 100,000 Monte Carlo simulations. The errors for fractional polarization and debiased polarized intensity are calculated with error propagation from intensity uncertainty in Stokes I, Q, and U. The NIR polarization null point is marked as a gray shadow. The position angle is the angle deviated from the galactic plane, and the positive value means the counterclockwise orientation from the galactic plane. The total intensity is normalized by the intensity of the center in each plane of 2307, 458, and 567 MJy/sr in the midplane, eastside, and westside about the galactic plane, respectively.} \label{fig:pol_asR}
\end{figure}

Figure \ref{fig:pol_asR} shows position angles with respect to the galactic plane \citep[22.9\degr;][]{2014A&A...565A...4H}, debiased fractional polarization, polarized intensity, and total intensity as a function of the distance from the center along the major axis of the galactic disk. All values are estimated within an area with a pixel size of 20.4\arcsec, the same as the right panel in Figure \ref{fig:polmap}. The total intensity panel shows three prominent peaks. One is at the nucleus, one is offset about +3.5 kpc to the northeast, and one is offset about -2.5 kpc to the southwest. In \citet{2020AJ....160..167J}, these regions offset from the nucleus were identified with possible spiral features seen in projection on the sky.

Position angles in the mid-plane are well aligned with the position angle of the galactic plane between -4 and +4 kpc. Although the debiased fractional polarization shown in Figure \ref{fig:pol_asR} is very small in the mid-plane, the S/N is high and the polarization angles are very well aligned with the disk plane, with one clear exception at +4.5 kpc. Position angles in pixels offset $20.4\arcsec$ to the west (blue line) and east (green line) of the galactic plane into the halo have greater errors. The east side has some regions with relatively coherent position angles. Position angles starting at -3 and going to +4 kpc on the east side trend positive (counterclockwise) as the distance along the disk from the center increases. This trend may be a part of an X-shaped field geometry often observed in radio wavelengths for edge-on galaxies \citep{2009RMxAC..36...25K}.

The second and third panel in Figure \ref{fig:pol_asR} is the debiased fractional polarization and polarized intensity plotted with respect to the distance from the center. The uncertainty in fractional polarization and polarized intensities at both ends of the galactic disk is larger because uncertainty in Stokes Q and U is higher in the outer disk than in the center. The uncertainty in Stokes parameters is roughly 2-3 times in the outer disk than in the center. Three locations with low polarization are seen in the galactic mid-plane (red line). The first one is at the center of the galactic plane. The other two are located near -3 and +4 kpc. The region near -3 kpc is close to a local total intensity peak in the southwest disk. The region near +4 kpc is just beyond the total intensity peak in the northeast disk. These intensity peaks along the disk from the center are likely the locations of star-forming regions, perhaps where the LOS is tangent to a spiral feature. The plausible causes of lower fractional polarization in the center and the intensity peaks will be examined later in this paper. 

In Figure \ref{fig:pol_asR}, the gray shadow in the northeast part of the disk depicts the location of a polarization null (in extinction) region identified at NIR wavelengths by \citet{2014ApJ...786...41M}. They found the fractional polarization dropped close to zero in this region. The FIR total intensity peak is about 0.5-1.0 kpc interior to the null point. We note that only at the NIR null point does the position angle in the disk significantly depart from the position angle of the plane (see the yellow box in Figure \ref{fig:polmap}). The vertical feature at the NIR null point is maintained up to $\sim$50$\arcsec$ off on either side of the galactic plane. This feature will be discussed in section \ref{dis:NIRnull}. The FIR fractional polarization is also low at this location, but the minimum FIR polarization is closer to the FIR intensity peak. The models from \citet{1997AJ....114.1405W} and \citet{2018ApJ...862...87S} that only consider optical/NIR polarization predict any null points in NGC891 would be due to the cancellation of polarization by mixing dichroic extinction and scattering. But, \citet{2014ApJ...786...41M} suggest that the polarization null point is not due to this cancellation effect as the null points are not symmetric. Also, our FIR observations are immune to the effect of scattered light in the beam, yet they clearly show lower polarization near the intensity peaks.

\subsection{Polarimetry outside the galactic disk} \label{subsec:offplane}

In Figure \ref{fig:polmap}, we find only a few polarization lines off the plane that meet our S/N cuts, and some are highly tilted from the galactic plane. Radio observations of NGC 891 show the presence of vertical magnetic fields in some regions in a halo of NGC 891 \citep{2020A&A...639A.112K}. It is hard to infer reliable FIR position angles within the small regions measured with the radio observations over the entire halo region due to our low S/N in fractional polarization at $154\micron$ outside the galactic disk. Instead, we select four regions where radio observations have well-detected polarization and compare the distributions of position angles observed in the FIR and radio polarimetry in each region. First, we will examine the position angle distributions in these regions and then discuss possible reasons for the low fractional polarization we measure off the plane.

\begin{figure*}[t]
\centering
\includegraphics[width=.8\linewidth]{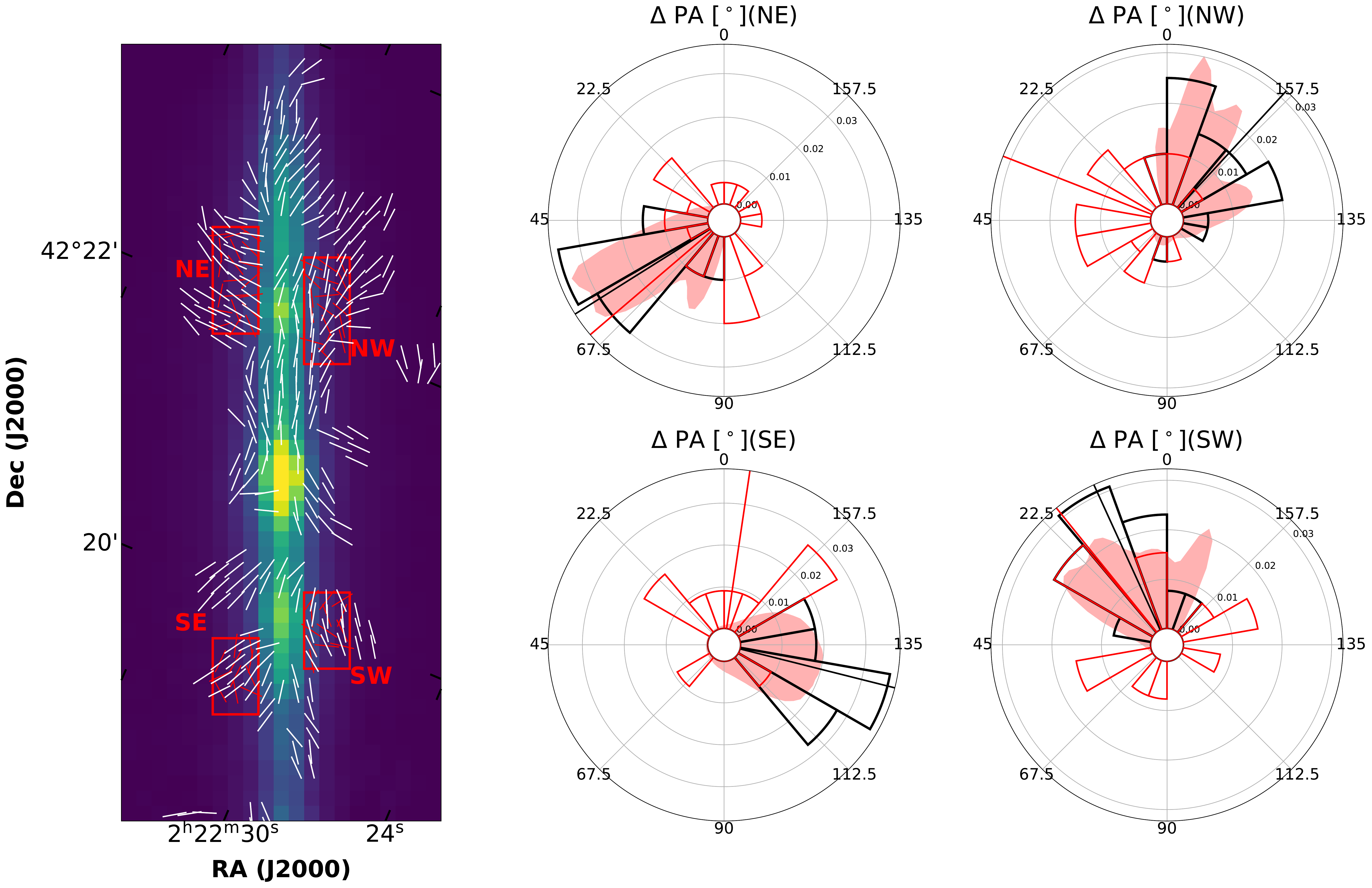}
\caption{The color map is the HAWC+ observation at $154\micron$ that the radio (white) and FIR (red) polarization lines are plotted over. The radio data is based on a polarization map with a 1.5$\arcsec$ grid with a beam size of 12$\arcsec$. The pixels with Stokes Q and U below 3 sigma are not provided for the radio data. Radio polarization lines nearest to the HAWC+ observation grid have been plotted and used. FIR data, which are in the red boxes and have the corresponding radio data, are only plotted and used here. The length of polarization lines is constant regardless of fractional polarization. Polar plots show the normalized distributions of position angles about the galactic plane, $\Delta PA$, within four regions indicated as red boxes over the color map. The selected area is $20.4\arcsec \times (34.0-47.6)\arcsec$. The black and red histograms are the normalized distributions of the position angles from radio and $154\micron$ observations. The expected distribution of our FIR observations (see the detail in the text) is plotted as the red shade. The black and red vertical lines indicate the circular mean value on the interval (-$\pi$/2, $\pi$/2), which is $\textrm{atan}2(\Sigma~\textrm{sin}2\theta, \Sigma~\textrm{cos}2\theta)/2$, of radio and FIR data, respectively.} \label{fig:reg4}
\end{figure*}

NGC891 has been observed at the wavelength of 5cm as a part of CHANG-ES \citep{2012AJ....144...43I}. The released data has been published in \citet{2015AJ....150...81W} and \citet{2020A&A...639A.112K}, and the polarization data has a grid resolution of 1.5$\arcsec$ and a beam size of 12$\arcsec$. We use the radio polarization lines nearest to the HAWC+ observation grid pixels, which are plotted in Figure \ref{fig:reg4}. Four rectangular boxes are overlaid on the map and labeled by their location. To the east of the galactic plane, radio observations reveal regions with a magnetic field mostly perpendicular to the galactic plane. The fields in the southwest region lie along the galactic plane. The northwest region has the magnetic field gradually rotating from mostly along to near perpendicular to the galactic plane. These structures were considered as evidence for an X-shaped structure in the lower angular resolution of $84\arcsec$ observations in \citet{2009RMxAC..36...25K}.

In Figure \ref{fig:reg4}, we show polar histograms of the position angles from each region marked on the map. The histograms are normalized so that the value of the area under the values sums to 1. A number density histogram of the polarization angles from radio and HAWC+ $154\micron$ observations are depicted with black and red lines, respectively. All of the data observed at $154\micron$ in a pixel size of $6.8\arcsec$ are included regardless of the S/N. We want to avoid using only a few pixels with a high S/N cut in debiased fractional polarization, which can lead to loss of position angle information. The FIR position angles appear to be more dispersed than those observed at radio wavelengths, especially in the northeast and southwest regions. Below we quantify this result.

It is possible that the observed $154\micron$ intensity in the halo is simply too faint for HAWC+ to detect several percent fractional polarization, given the observed errors in Stokes I, Q, and U. If HAWC+ had too low sensitivity, we would expect a random distribution in position angles due to noise alone. We can test this possibility by computing an expected distribution of position angles based on the observed Stokes intensity strength and errors in the actual $154\micron$ data using an assumed value for the expected fractional polarization and position angle. The expected distributions are simulated by computing a range in values using a Gaussian distribution of these (observed) errors about a mean for the expected Stokes Q and U intensities. The expected Stokes Q and U intensities are computed assuming an intrinsic polarization angle and an intrinsic fractional polarization. The polarization angles are assumed to be coincident with the radio polarization angles. We set fractional polarization at 9\%, which is the expected maximum (see section \ref{subsec:models_param}). We note that for the Milky Way, the fractional interstellar polarization is maximum both in the optical extinction \citep{2018A&A...616A..52S} and in emission \citep{2020A&A...641A..12P} at high latitudes and low extinction lines of sight. 

The simulations are run 100,000 times in each pixel given the intrinsic polarization angle, the intrinsic fractional polarization, and the observed Stokes Q and U uncertainties. The normalized distribution of all simulated position angles represents the expected distribution of position angles within a selected region, and the normalized distribution is shown as the red-shaded histogram in Figure \ref{fig:reg4}. These computed distributions are very tight and indicate that $if$ the magnetic field was in the plane of the sky and relatively uniform in the halo of NGC 891 $and$ the grains were emitting polarized light with reasonable efficiency, we would have clearly detected far more polarization lines than we did. In other words, the observed weak FIR polarization signal and chaotic position angle distribution we find in the halo is likely $not$ due to the observed intensity being below our detection limit, but is a feature of the galaxy.

To statistically quantify the randomness of the distribution in observed polarization angles at $154\micron$ and the mean difference between $154\micron$ and radio polarization angles in Figure \ref{fig:reg4}, we ran the Rayleigh test and v-test in Astropy package \citep{2013A&A...558A..33A,2018AJ....156..123A}. A small p-value from the Rayleigh test supports the non-uniformity of the 154$\micron$ position angle distribution. The p-values for the Rayleigh test are 0.519, 0.002, 0.060, and 0.574 in the northeast, northwest, southeast, and southwest region, respectively. The v-test confirms whether the data is distributed uniformly or assumed to have the same mean value as the mean position angle from radio observations. A high p-value can be due to uniformity or a difference between the mean values of $154\micron$ position angles and the radio position angles. The p-values from the v-test are 0.128, 0.894, 0.590, and 0.153 in the northeast, northwest, southeast, and southwest region, respectively. The two test results indicate the following: 1) the position angles in the northeast, the southwest, and (probably) the southeast regions have a uniform distribution, which may be caused by a reduction in fractional polarization (to below our detection limit) relative to the maximum expectation or may be due to an actual random magnetic field geometry; And, 2) the northwest region has position angles with a nonuniform distribution, and the mean orientations of the $154\micron$ and radio polarizations are statistically different.

To see how low a fractional polarization is required for the uniform distribution of position angles shown in the northeast and southwest regions, we ran the simulations described above, but varied the maximum fraction polarization (initially 9\%) and then performed the same Rayleigh test. The 7 of 10 simulations in the northeast and southwest regions with the intrinsic polarization of 1.5\% give the p-value larger than 0.05, implying we cannot reject the hypothesis of uniformity. Therefore, assuming the magnetic field lies in the plane of the sky, the intrinsic polarization would have to be less than 1.5\% to explain chaotic position angle distributions. Our models will show that when the magnetic field is parallel to the galactic plane and has a spiral pattern, the expected mean values of the fractional polarization in the selected regions should be greater than 3\% if this geometry is maintained up into the halo. The low fractional polarization we measure in the halo may be caused by a lower intrinsic polarization for the dust grains, the effect of turbulence, and/or a mixture of planar and vertical fields that partially cancel the net polarization. We will discuss this further in section \ref{dis:geo_off}.

\section{Model Polarimetry in the galactic disk} \label{result:model_result}

Our results in section \ref{sec:obs_result} have affirmed that the magnetic field near the galactic plane projected on the plane of the sky lies close to the position angle of the plane, and there are clear locations with low fractional polarization along the galactic disk (Figure \ref{fig:polmap} and \ref{fig:pol_asR}). Polarimetry studies in the Milky Way indicate the fractional polarization is affected primarily by the changes in the magnetic field geometry along the LOS  rather than grain properties \citep[e.g.,][]{1992ApJ...389..602J, 2020A&A...641A..12P}. Problems with grain alignment in dense molecular cloud cores \citep{2019ApJ...882..113S} are unlikely to be a factor due to the small filling factor of these regions in our HAWC+ beam \citep{2020ApJ...888...66L, 2020AJ....160..167J}. In this section, we will develop a set of simple models that assume a planar magnetic field in the disk of the galaxy for comparison with the observations.

\subsection{Model descriptions} \label{subsec:models_param}

A synthetic galaxy is modelled in a volume of 14.1 $\times$ 14.1 $\times$ 3.5 kpc$^3$ with a grid cell size of 9.2 pc, much smaller than our beam. Making a polarization map from this synthetic galaxy is performed in three steps. First, we create a dust density distribution that resembles the FIR maps commonly seen in face-on disk galaxies. Appendix \ref{app:density} explains the details of the dust density distribution used in our models.

Next, we place the magnetic field geometry threading the dusty disk. Magnetic fields in galaxies are often characterized by consisting of an ordered component and an isotropic random component associated with turbulence \citep[following the terminology in][]{2013pss5.book..641B}. The ordered component of the magnetic field is sometimes referred to as the constant, or large-scale component in previous studies \citep[e.g.,][]{1992ApJ...389..602J, 2009ApJ...696..567H}. For the ordered magnetic field component, we assume all ordered magnetic fields lie in the galactic plane and rely on known observations of the relation between the ordered magnetic fields and disk galaxy morphology (see Appendix \ref{app:orderedB}). We use two simple geometries for the dust and the ordered magnetic field component, a simple spiral with a constant pitch angle and a barred spiral.

Note that all the parameters for modeling the density distribution and the ordered magnetic field are adjusted independently in the northeast and southwest disk. The bright spots off the nucleus in NGC891 are not symmetric, as confirmed in many studies \citep[e.g.,][]{1980MNRAS.193..313B} and shown in Figure \ref{fig:pol_asR}. The model dust distribution and the ordered magnetic field geometry in a spiral and a barred spiral galaxy are depicted in the left panels in Figure \ref{fig:model_param}. 

In addition to the ordered magnetic field, an isotropic random component is assigned to each `turbulence cell', which is larger than the model grid size. That is, several grid cells share the same turbulent magnetic field geometry within a turbulent cell. The method used to divide up the model galaxy into turbulence cells is given in Appendix \ref{app:turbcells}. Each turbulent cell is adjusted to have a total neutral and molecular hydrogen number close to a chosen value, so that denser regions have smaller turbulent cells. Thus, higher column depth lines of sight will pass through more turbulent cells \citep[e.g.,][]{1992ApJ...389..602J, 2020A&A...641A..12P}. The turbulent component is taken to be isotropic with a Gaussian distribution, similar to the model by \citet{1991ApJ...373..509M}. The Gaussian distribution is scaled by the ratio of the turbulent ($B_t$) to the ordered ($B_0$) magnetic field strengths, the parameter $B_t / B_0$. Here, $B_t / B_0$ is the ratio of magnetic field strengths. Note that the Gaussian dispersion of the random component along one independent axis, which is referred to as $\sigma_B$ in \citet{1992ApJ...389..602J}, is $B_t / \sqrt{3}$ in this work.

Lastly, the polarized emission is integrated through the disk from back to front using the equations of transfer in the Appendix of \citet{2014ApJ...797...74D}. The FIR radiation emitted from each grid cell is calculated given the optical depth and black body temperature, where we assume the dust temperature of 23K everywhere. The dust temperature computed from the FIR colors as viewed edge-on does not change much with location \citep{2014A&A...565A...4H}, and we have no information on how dust temperature varies in NGC 891 due to lack of a face-on view. The optical depth is calculated from $\rho\textrm{(HI+H}_2)$ with the parameters given in Appendix \ref{app:density}. Finally, the integrated emission for the edge-on view of a galaxy is convolved with a Gaussian kernel with a scale of $13.6\arcsec$ (the HAWC+ FWHM beam) and regridded to a pixel size of $20.4\arcsec$, the same spatial grid used in Figure \ref{fig:pol_asR}. Based on the intensity errors derived from the observations in Stokes I, Q, and U, we make model images in Stokes I, Q, and U that include the effects of observational error. By including observational error in the model, we can make comparisons with the observed fractional polarization, uncorrected for bias.

Many realizations of the model are computed with the random component in each turbulent cell uncorrelated with other turbulent cells. The model output for the net polarization along each LOS contains a median value and a dispersion in both the fractional polarization and the position angle due to variations in the contribution of the random component in the individual realizations $and$ the contribution of observational error. As we will see, the relatively smoothly varying fractional polarization and position angle in the data for the disk will require large numbers of turbulent cells for the model predictions to make sense. With only a few cells, the model will predict a much wider distribution in fractional polarization and position angle than is observed (see the left panel of Figure \ref{fig:turb_model}).

We tested models by varying the geometry of the dust density, the contribution of the turbulent magnetic field, a mixture of magnetic field orientations, and the maximum fractional polarization of dust grains. We chose to explore two different types of galaxy morphology: a simple spiral and a barred spiral galaxy. The contribution of the turbulent magnetic field is determined by two parameters: one is the number of turbulence cells in the model galaxy, and the other is the ratio $B_t / B_0$, the relative strength of the turbulent and ordered components. The number of turbulence cells is determined by a threshold number parameter, $\mathcal{N}_{th}$, defined in Appendix \ref{app:turbcells}. The smaller this number is, the smaller size and hence the greater number of turbulent cells. The effects of changing this value are described in Appendix \ref{app:turbcells}. The initial results shown in section \ref{subsec:models_comp} are modelled with $\mathcal{N}_{th}=10^{63}$, which roughly corresponds to A$_V \sim 1$ for a turbulent cell, the value found by \cite{1992ApJ...389..602J}. We first assume $B_t / B_0 = 1$, which works well for the diffuse interstellar medium in the Milky Way \citep{1992ApJ...389..602J, 2008A&A...490.1093M}, but consider a larger value as well. Values for $B_t / B_0$ higher than 1 have been measured in external spiral galaxies with radio observations \citep{2019Galax...8....4B}. We will find that increasing $B_t / B_0$ above this Milky Way value will require decreasing $\mathcal{N}_{th}$ (increasing the number of turbulent cells) for the model to match the observations.

We initially assumed the ordered magnetic field component is strictly parallel to the galactic plane, then we considered the effects of having some fraction of the gas containing a vertical field, perpendicular to the plane. For simplicity, this fraction was assumed to be uniform throughout the galaxy. Finally, the maximum fractional polarization of thermal dust grain emission, $p_{max}$, which is the fractional polarization when the magnetic field is perpendicular to the LOS, was allowed to vary. We initially use $p_{max}$ = 9\%, taken from  \citet{1995ApJ...450..663H}. In terms of model output, adding a vertical component and lowering $p_{max}$ have nearly the same result. The emission is optically thin, and adding a vertical component to the parallel field reduces the net polarization per unit optical depth, as does reducing $p_{max}$. We kept these two effects separate in order to individually quantify each parameter.

\subsection{Comparisons with observations} \label{subsec:models_comp}

\begin{figure*}[t]
\centering
\includegraphics[width=.8\linewidth]{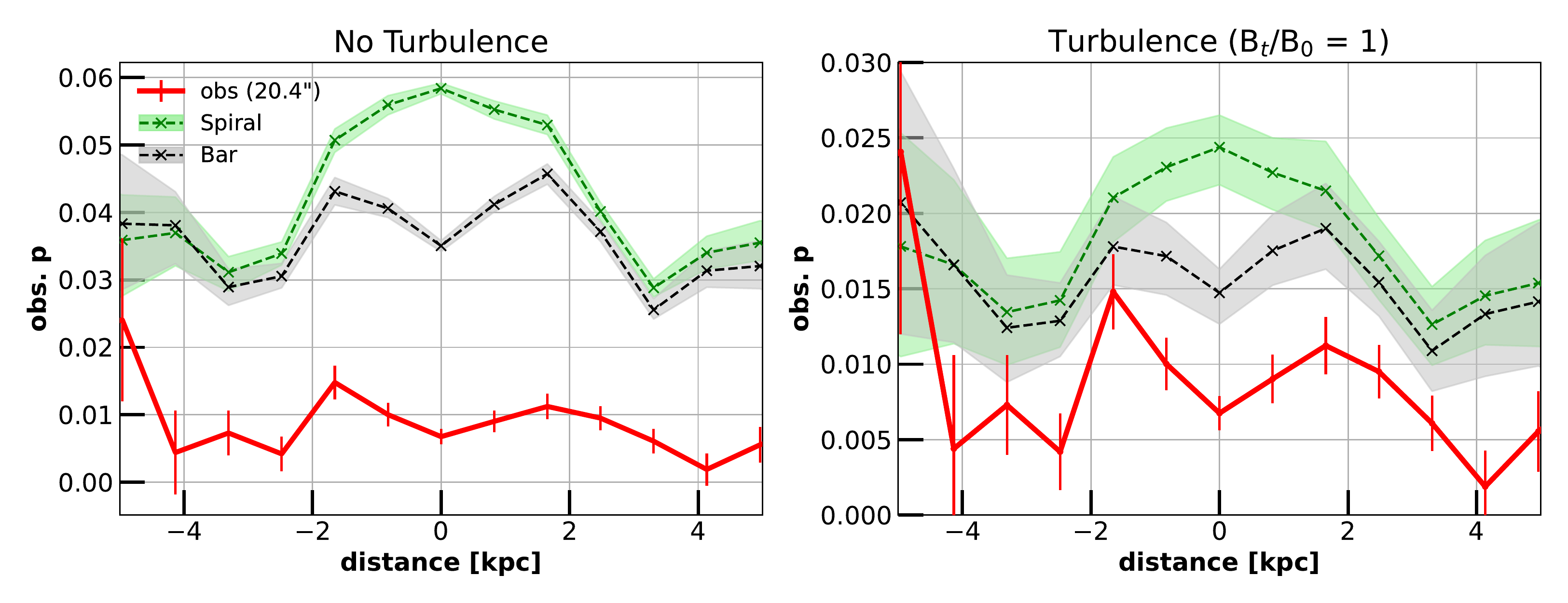}
\caption{Fractional polarization of a spiral (green) and a barred spiral (black) model galaxy with an edge-on view. No turbulent magnetic fields exist in the left panel. The ratio of turbulent to ordered magnetic fields, $B_t / B_0$, is added as 1.0 in the right panel. The transparent colored regions surrounding the model median values represent a dispersion width of 60\% in 500 simulations. Note that observational errors are included in the model results, hence the greater dispersion at larger distances from the center where the galaxy is fainter, even in the case of no turbulence. The red line shows the observed fractional polarization in the galactic mid-plane without any S/N cuts, and not debiased.} \label{fig:model}
\end{figure*}

The comparison between synthetic models and observations is performed only in the galactic midplane. Our observations show the magnetic field geometry is more complicated away from the galactic disk. We first compare the observations with a synthetic model for a simple spiral galaxy. The spiral galaxy model shown in Figure \ref{fig:model} reproduces the decrease of fractional polarization near -2.5 and +3.5 kpc, where the FIR emission integrated along the LOS has peaked. This decline is due to the LOS being close to a tangent to the spiral arm and consequently parallel to the magnetic field orientation. An important feature of the observations shown in Figure \ref{fig:pol_asR} is the depression of fractional polarization at the center of the galaxy. Our synthetic spiral galaxy differs from the observation results, and it shows the greatest fractional polarization towards the center of the galaxy. This is due to the model field lines crossing nearly perpendicular to the LOS along the path through the galactic center. This implies that a simple spiral geometry (or any circular structure) cannot produce a model polarization that matches the data in the center of NGC 891. 

We should mention that pitch angles in galaxies can vary with a radius and that a magnetic pitch angle larger than $50\degr$ can create a central dip in fractional polarization, but with adverse effects outside the central region. Using two pitch angles in the model is beyond what we can constrain with the edge-on view. The presence of spiral arms in NGC891 has been proposed based on the color differences \citep{1981A&A....95..116V} and asymmetry in H$\alpha$ emission \citep{2007A&A...471L...1K}. However, the presence or absence of a bar structure or a spiral structure in the center is not ruled out.

Using our model with a bar structure, the model fractional polarization in the center decreases enough to match the observations, as shown in Figure \ref{fig:model}. Note that the decline in fractional polarization in the center depends on the position angle of the bar axis in the model with respect to a LOS, and the assumed geometry of the magnetic field in the bar is based on radio observations of face-on barred spirals (see Appendix \ref{app:orderedB}). Our simple modeling shows that the observed magnetic field geometry of NGC 891 can not be fit using a simple spiral pattern but requires a geometry similar to that found in a barred galaxy. Although adding a tilted bar in the center can reproduce the lower fractional polarization in the center, the magnitude of the polarization in our model without turbulence significantly exceeds the observational results over the entire disk. 

In the right panel, a turbulent magnetic field is added to the model with $B_t$/$B_0 = 1$. As expected, the presence of turbulent magnetic fields lowers the net fractional polarization compared to the case without turbulence. However, the model polarization is still too high. Note, the model with a turbulent component predicts a larger dispersion in fractional polarization compared to the effects of observational error alone. The model dispersion can initially be lowered by increasing the number of turbulent cells (decreasing $\mathcal{N}_{th}$), but, increasing the number of turbulent cells beyond 8 or so in a LOS does not continue to significantly lower the net fractional polarization. This is because the net fractional polarization effectively saturates with an increasing number of turbulent cells beyond a few \citep[see Appendix \ref{app:turbcells}; Figure 12 in][]{2016A&A...596A.105P} until optical depth effects become important.

Increasing the number of turbulence cells $does$ affect the predicted dispersion of fractional polarizations and position angles, even though the median values of simulated fractional polarization tend to saturate (left panel of Figure \ref{fig:turb_model}). The fact that the observed fractional polarizations and position angles in the disk vary little with location (see Figure \ref{fig:pol_asR}) and agree with the median values of the simulation implies that the predicted dispersion of the polarization and position angles in the simulation must be small. There must be a large enough number of turbulence cells to average out the effect of randomness on the net position angle for any LOS through the disk plane.

\begin{figure}[t]
\centering
\includegraphics[width=.8\linewidth]{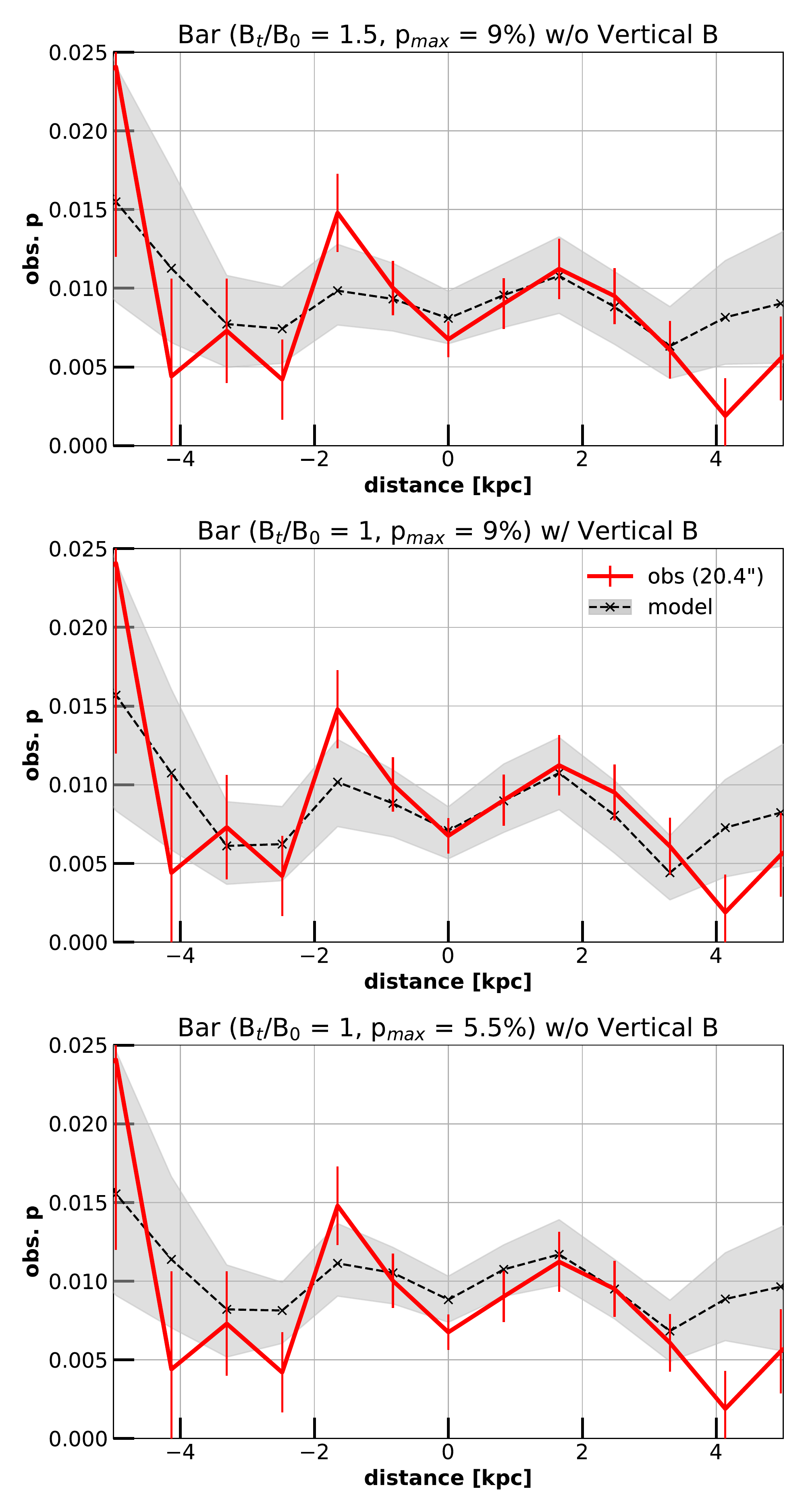}
\caption{Fractional polarization from the models in comparison with observations. These results are from a synthetic barred spiral galaxy. Different panels vary $B_t/B_0$, the presence of vertical fields, and $p_{max}$. A threshold value for turbulence cells, $\mathcal{N}_{th}$, for the modeling is used as $10^{63}$, but $5\times 10^{62}$ for the model with $B_t/B_0$ of 1.5. The black lines and grey shades represent the median value and the distribution of 60\% in 500 simulations.} \label{fig:model_3}
\end{figure}

As seen in Figure \ref{fig:model}, the model still predicts a higher fractional polarization than observed, even with the addition of a turbulent component to the field. Additional factors in reducing the fractional polarization are necessary for our models to match the very low fractional polarization in the data. We can increase the effect of a turbulent component on the fractional polarization by increasing $B_t / B_0$. It is possible the relative strength of the turbulent component in NGC891 is greater than in the Milky Way due to the enhanced star formation rate of a factor of 2-3 compared to the Milky Way \citep{2004A&A...414...45P, 2015ApJ...806...96L}. The top part in Figure \ref{fig:model_3} shows that the fractional polarization in the model becomes comparable to the observations when $B_t / B_0$ is increased to 1.5. Such enhanced $B_t / B_0$ has been suggested in M51 by \citet{2013ApJ...766...49H} using a structure function analysis. To keep the model dispersion in fractional polarization and position angle concordant with the observations, it was necessary to reduce the value of $\mathcal{N}_{th}$ by a factor of two, increasing the number of turbulent cells along any given LOS. Otherwise, the model predicts more dispersion in fractional polarization and position angle than observed.

\begin{figure*}[t]
\centering
\includegraphics[width=.8\linewidth]{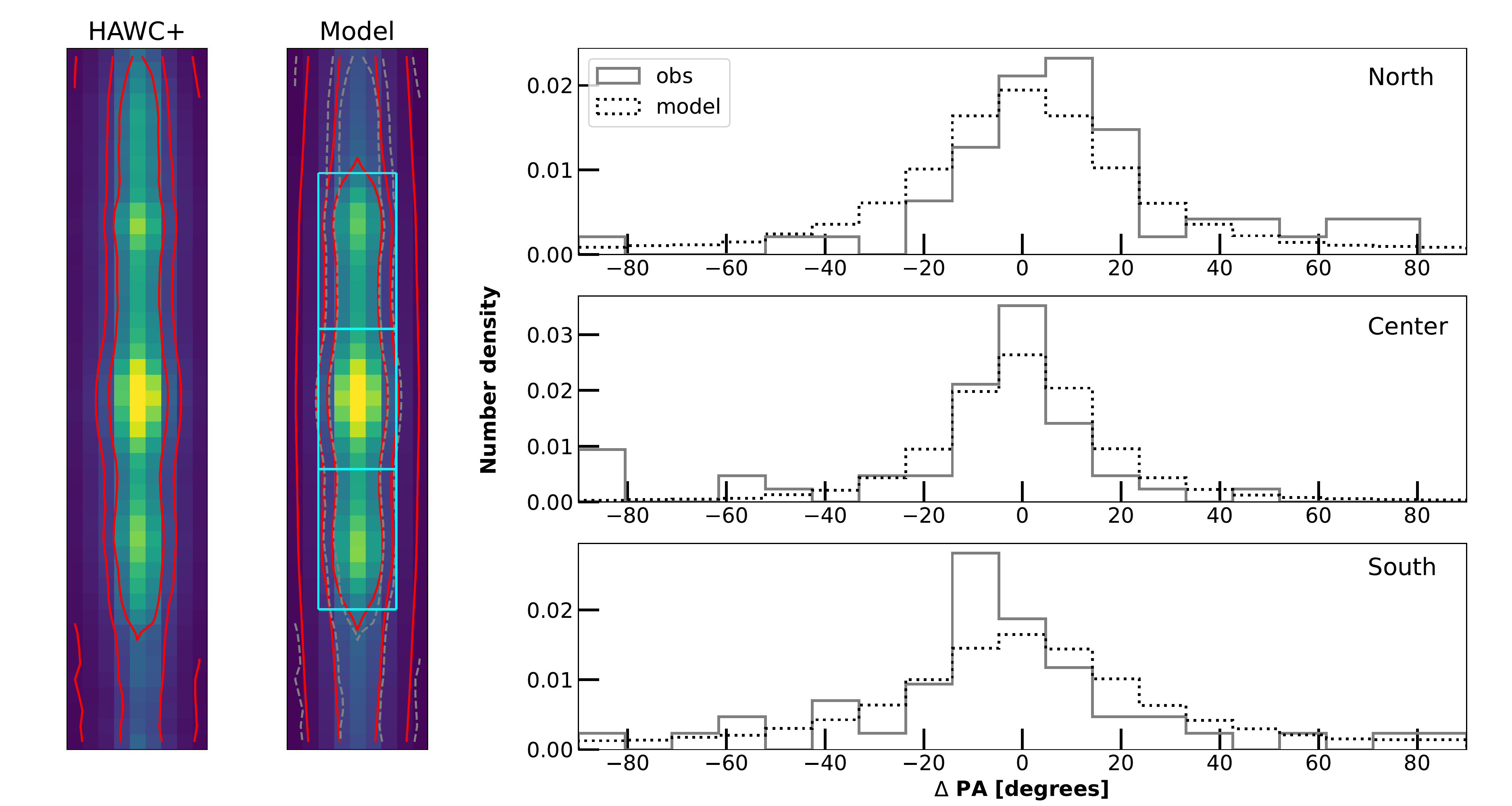}
\caption{Left: HAWC+ total intensity observation at 154\micron, Middle: Our model with a bar structure on an edge-on view. The red contours are 100, 500, and 1000 MJy/sr in both HAWC+ observation and the model. The gray dashed contours in the middle are the same as the contours in the HAWC+ observation. Right: Number density distribution of position angles within the north, the center, and the south area. Each region is marked as cyan boxes in the middle. The position angle is measured from the galactic plane, and the counter-clockwise direction is a positive value. Solid lines are the position angles of pixels within each region from the HAWC+ observation. Dashed lines are the density distribution derived from 500 synthetic images. The model results are from the model with $B_t/B_0 = 1.5$, which shown in the top panel in Figure \ref{fig:model_3}.} \label{fig:NcenS}
\end{figure*}

A comparison between the model position angle distribution with an enhanced $B_t / B_0$ and the observations is made in Figure \ref{fig:NcenS}, where we have divided up the disk into three regions as shown in the model intensity map. Note that 0$\degr$ is the model disk plane, corresponding to $22.9\degr$ on the sky. The mean values and the dispersion of the position angle distribution for both the model (which includes observational error and the effects of turbulence) and the observations are very similar. We estimated the circular mean, devised for cyclic quantities, and the Kuiper test of the observation data with Astropy package \citep{2013A&A...558A..33A,2018AJ....156..123A}. The mean values in the north, center, and south, are 8.9$\degr$, -5.4$\degr$, and -6.4$\degr$ respectively. The p-values between the modeled and observed position angles based on the Kuiper test are 0.16, 0.11, and 0.21 in the north, center, and south. According to the p-values, there is no clear statistical difference between the observations and our synthetic models. This implies that our assumption that the ordered component of the magnetic field lies in the galactic plane is realistic and that a significant number of turbulent cells is required to explain the narrow observed position angle distribution. Note that the small angle offsets seen in the FIR are not statistically different from zero and are coincident with neither radio data nor the observations at NIR wavelengths \citep[see Figure 8 in][]{2014ApJ...786...41M}.

Another possible mechanism for lowering the fractional polarization is the presence of vertical magnetic fields, which effectively cancel a part of the polarization due to dust grains aligned with the field in the plane. Radio observations reveal several galaxies with vertical magnetic fields in their halo \citep{2015A&ARv..24....4B}. But, FIR observations have found  magnetic fields that are clearly perpendicular to the galactic plane only in M82 \citep{2019ApJ...870L...9J} and NGC 2146 \citep{2022ApJ...936...92L}, which are strong starburst galaxies, and Cen A, which has an active AGN \citep{2021NatAs...5..604L}. The low NIR polarization in NGC 891 has been explained by the assumption that about 1/3 of the disk gas contains a vertical magnetic field \citep{1997AJ....114.1393J}. In the middle of Figure \ref{fig:model_3}, we assume that 15 percent of dust contains magnetic fields vertical to a galactic plane. Our model result shows that this is enough to lower the model fractional polarization to match the observations. This result looks promising, and section \ref{subsec:offplane} shows the possibility of the mixture of planar and vertical magnetic fields. However, we have assumed in our model that the vertical fields would be mixed uniformly throughout the disk, which is unlikely. Adding extra parameters to describe specific locations for a vertical field component would over-complicate the model.

Lastly, we adjusted the maximum fractional polarization, $p_{max}$ in our models. The initial value of 9\% is based on observations of other galaxies with HAWC+ \citep{2020AJ....160..167J} and is similar to the maximum values seen in star-forming regions in the Milky Way \citep{1995ApJ...450..663H} and a value from dust models \citep{2018A&A...610A..16G}. By using $p_{max}=5.5\%$, we can lower the model polarization to match the observations. The maximum fractional polarization in dust emission is primarily determined by the physical characteristics of the dust. If a reduced $p_{max}$ is necessary for our model to match the observations, NGC 891 must have a significantly different dust population compared to the Milky Way and other external galaxies. Without corroborating evidence for a very different dust population in NGC 891, we think this seems unlikely.

Our models with a barred spiral galaxy using enhanced turbulence in the magnetic field, a mixture of vertical and parallel fields, or lower polarizing power of the dust grains, reasonably fit observations within the galactic disk. The models reproduce the valleys of fractional polarization in the center and two bright spots off the center along the galactic disk, although the fractional polarization in the two bright spots is a little higher in our model. Given it is unlikely that the dust in NGC 891 has significantly different physical properties than in other normal disk galaxies and the lack of any clear evidence for vertical fields in the entire disk, we find that enhanced turbulence with short decorrelation lengths best describes our FIR polarimetry observations.

Although our simple model assumes the constant ratio of turbulent to ordered magnetic fields, one alternative model method is creating a ratio of turbulent to ordered magnetic fields depending on gas density, $B_t/B_0 \propto \rho^{0.5}$. The small-scale dynamo predicts equipartition between magnetic and kinetic energy densities, $B_{tot} (=\sqrt{B_{0}^2 + B_{t}^2}) \propto \rho^{0.5}$, and almost uniform ordered magnetic energy density has been observed, for example, in IC 342 \citep[see][]{2015A&ARv..24....4B}. A higher ratio of turbulent to ordered magnetic fields in a higher density region can bring about reduced polarization in the center of a model spiral galaxy. We tested this idea by allowing $B_t/B_0 \propto \rho^{0.5}$ with cell volume density. We found that values for $B_t/B_0$ large enough to match the overall low fractional polarization in the galaxy cannot reproduce the very low polarization near spiral arms, where the model has the ordered field along the LOS. Because of enhanced turbulence in the spiral arms (which are denser regions), the magnetic fields within spiral arms no longer strongly align with the arm direction down our LOS. It is hard to investigate these constraints with the limitations of an edge-on view and our spatial resolution. Some dependence of turbulence can not be ruled out (see section \ref{dis:geo_sp}), and the variation of the turbulent component with volume density should be explored with face-on galaxies observations in the future.

\section{Discussion} \label{sec:discussion}

\subsection{Center of the galaxy} \label{dis:geo_cen}

The fractional polarization of our simple spiral model galaxy is in disagreement with the very low FIR polarization along the LOS through the center. The simple spiral model predicts a much higher fractional polarization due to the ordered component of the magnetic field being nearly perpendicular to our LOS. The presence of a bar structure in NGC891 is capable of reproducing the central drop-in fractional polarization seen in the observations. In our bar structure, the magnetic field is aligned along the bar axis \citep{2005A&A...444..739B}. Tilting the bar relative to the plane of the sky can cause a reduction in fractional polarization along the LOS through the center. Several studies have suspected the existence of a bar in NGC891. Gas kinematics estimated with CO observations, for example, can be reproduced by a model with the gas flow driven by a bar \citep{1995A&A...299..657G}. NIR imaging observations suggest a bar-like structure that is thick and truncated at $\sim$ 3kpc \citep{2013ApJ...773...45S}.

\subsection{The effect of spiral arms} \label{dis:geo_sp}

In the disk of NGC 891, two positions on either side of the center have relatively lower fractional polarization at FIR wavelengths. Our modeled galaxies in Figure \ref{fig:model} and \ref{fig:model_3} confirm that looking down spiral arms in an edge-on view has low fractional polarization at FIR wavelengths. The observed low polarization locations on either side of the nucleus in Figure \ref{fig:pol_asR} are likely due to the magnetic fields within spiral arms, which is parallel to the LOS at these locations.

Comparing our observations with the model results, the observed regions with the lowest fractional polarization do not line up exactly with the peak of intensity. First, in real galaxies, the pitch angles of magnetic fields and spiral arms are not exactly coincident, and both pitch angles also can vary with location, unlike our model assumption \citep{2015ApJ...799...35V}. The somewhat different magnetic pitch angles between the arm and inter-arm regions, which may be caused by the compression, have been observed in M51 \citep{2011MNRAS.412.2396F, 2021ApJ...921..128B}. Also, magnetic arms with highly ordered magnetic fields in the inter-arm regions can affect the observations. The magnetic arms have been found in most spiral galaxies, although their origin is not well understood \citep{2015A&ARv..24....4B}. It is probable that these complexities in real spiral galaxies can cause the location of the low polarization points to not exactly line up with the intensity peaks, but we find that regions with a LOS preferentially down the LOS best explain the low polarization regions. 

The FIR observations still show slightly lower fractional polarization compared to our simple models in the regions with fractional polarization minima along the galactic plane. Perhaps enhanced star formation in the spiral arms, in general, may create local blowouts and bubbles smaller than our beam, drag the magnetic field vertically up into the halo, and lower the net fractional polarization. It is also possible that an even greater turbulent component in the spiral arms further reduces the fractional polarization. It was found that the contribution of the turbulent magnetic field strength is more significant in the arms than that in inter-arm regions of a face-on galaxy, M51 \citep{2011MNRAS.412.2396F, 2020A&A...642A.118K}. This suggests some dependence of the turbulent component on volume density could be a factor (see section \ref{subsec:models_comp}).

The deviation between the observed regions with the lowest fractional polarization and the total intensity peaks is larger in the northern disk than in the southern disk. NIR observations in \citet{2014ApJ...786...41M} also found a polarization null point located at $\sim$4.5kpc in the northern disk. Our FIR lowest polarization point is between the peak of FIR total emission ($\sim$3.5 kpc from the center) and the NIR polarization null point. We will discuss the NIR polarization null point in the following subsection.

\subsection{Vertical feature at the NIR polarization null point} \label{dis:NIRnull}

Most FIR polarization lines in the disk align well with the galactic plane. But, we found one point where a significant departure to the planar field in the disk is apparent in the northern disk at a distance of $\sim$4.5 kpc ($\sim 110\arcsec$) from the center. The FIR position angle there is $\sim 90\degr$ to the disk, and this feature stretches off the plane on both sides into the halo, suggesting some sort of blowout (right hand panel, Figure \ref{fig:polmap}). We note that near this location, there is sparse synchrotron polarization (Figure \ref{fig:reg4}), probably due to Faraday depolarization across this complicated region. The synchrotron polarization that is measured is likely coming mostly from the front side.

This location in the northern disk is where a NIR polarization null point in NGC 891 is found by \citet{2014ApJ...786...41M} and where the observed $154\micron$ total intensity actually remains high compared to our model (which does not reproduce the excess emission in the north, see Figure \ref{fig:density_model}). In our FIR observations (Figure \ref{fig:pol_asR}), the lowest fractional polarization in the northern disk is at $\sim 4$ kpc, between the $154\micron$ intensity peak ($\sim$3.5 kpc from the center) and the NIR polarization null point ($\sim$4.5 kpc). Polarimetry at NIR wavelengths can not penetrate into the disk as deep as our optically thin FIR observations, and the NIR polarization null point is likely to be due to a region closer to the near side of the galaxy. The lowest FIR fractional polarization location is likely due to a combination of spiral arm geometry, which cannot be seen at NIR wavelengths, and a blowout causing a mixture of vertical and horizontal components to the NIR polarization creating the NIR null point.

The dust temperature has a distinct peak at this location in the disk \citep[see Figure 3 in ][]{2014A&A...565A...4H}. \citet{1997AJ....114.2463H} also found several dust features in optical images, which extend up to z$\sim$800pc ($\sim20\arcsec$) above and below the disk. The dust feature at the NIR null point is also associated with bright optical emission from ionized gas, which they associate with enhanced star formation in the disk causing a blowout. This phenomenon could strongly alter the ordered magnetic field geometry due to the creation of dusty pillars stretching up from the disk into the halo \citep[e.g.,][]{1998LNP...506..229H, 2018A&A...611L...5A}. An interaction with UGC 1807, a nearby companion \citep{2008MNRAS.388..697M}, could cause an increase in the star formation rate in the northern disk, with a blowout taking place at the NIR null point.

The radio synchrotron emission does show some hints near this location. Since the gas volume sampled by dust emission may contain significantly different magnetic field orientations compared to that measured in the radio, we can only tentatively associate the synchrotron polarization lines with a blowout from the disk. Just how strong a contribution the dusty regions in the halo make to the overall synchrotron emission is not clear.

\subsection{Fields outside the galactic disk} \label{dis:geo_off}

An X-shaped magnetic field geometry has been observed in the halo of edge-on galaxies in radio synchrotron observations \citep[e.g.,][]{1994A&A...284..777G, 2009RMxAC..36...25K, 2020A&A...639A.112K}. There are several studies that try to explain this feature theoretically. \citet{1993A&A...271...36B} suggest that dynamo models with a galactic wind drag magnetic fields from a galactic disk to a halo. \citet{2018MNRAS.477.3539N} propose an initial poloidal field evolving by shear effects between the warm interstellar medium and the halo. However, the mechanism forming the observed X-shape is not considered to be well understood yet \citep{2019Galax...7...36M}.

Early radio observations suggested NGC 891 has an X-shaped field geometry in the halo \citep{1994A&A...284..777G, 2009RMxAC..36...25K}. Recent radio observations show a more complicated magnetic field geometry in the halo, with some vertical fields in a few patchy regions \citep{2020A&A...639A.112K}. Optical polarimetry from \citet{1996MNRAS.278..519S} suggests the presence of vertical fields in the halo of NGC891. However, sensitive NIR polarimetry finds no evidence for this \citep{2014ApJ...786...41M} and contamination by scattering at optical wavelengths is likely \citep[see the discussion in][]{2020AJ....160..167J}, making it difficult to assess the optical polarimetry. 

In the comparison of $154\micron$ and at radio polarizations in Figure \ref{fig:reg4}, we see that the position angles in our FIR polarimetry often have an orientation inconsistent with the mean position angle of radio polarizations, or show no statistically significant mean direction. The inconsistency of position angles means that the gas volume sampled by the dust emission contains significantly different magnetic field orientations compared to that seen in the radio. It is probable that the dusty regions in the halo are only weak contributors to the overall synchrotron emission, but constitute the major fraction of the FIR emission. Position angles at $154\micron$ seem to be randomly distributed in most regions, indicating either a very scrambled field geometry unlike the synchrotron polarimetry and/or some sort of cancellation of polarized emission causing the weaker fractional polarization in the halo than expected. It is very unlikely NGC 891 would have anomalously low intrinsic dust grain polarization in the halo unless the dust population is very different than in the Milky Way.

We would generally expect less turbulence in these halo filaments due to the low optical depth. In the Milky Way, high latitude and low extinction LOSs produce the maximum observed interstellar polarization in both extinction and emission \citep{2018A&A...616A..52S, 2020A&A...641A..12P}. Optical images of NGC 891 confirm the presence of many dusty filaments that extend vertically above and below the disk \citep{1997AJ....114.2463H}. These dust features may contain vertical magnetic fields, as suggested in \citet{1994AJ....108.2102S}. Therefore, we suggest that the low polarization signal at FIR wavelengths in the halo of NGC 891 is attributed by a mix of vertical and horizontal field geometries that causes partial cancellation between vertically and horizontally polarized dust emission.

There are an HI filament that stretches outward the northwest \citep{2007AJ....134.1019O} and a FIR spur in the southeast of NGC891 \citep{2021MNRAS.502..969Y} that may be due to tidal interaction or some sort of feedback process. We may expect some signature of these features in our $154\micron$ polarimetry, but there is no clear trend.

\subsection{Degree of polarization} \label{dis:fracp}

As shown in Figure \ref{fig:model}, our simple model using only the ordered component of the magnetic field predicts significantly higher fractional polarization than observed. To lower the model fractional polarization, we explored three model inputs: 1) Adding a turbulent component to the magnetic field, 2) Adding some fraction of the gas with a vertical magnetic field everywhere in the model galaxy, and 3) Reducing the value for the maximum fractional polarization that the dust can produce.

As shown in Figure \ref{fig:model}, adding turbulence magnetic fields in addition to ordered magnetic fields reduces the fractional polarization, as expected. The ratio of $B_t / B_0= 1.5$, which is larger than $B_t / B_0 = 1$ based on observations in the Milky Way \citep{1992ApJ...389..602J}, is enough in our model to lower the model fractional polarization to levels comparable to the observed values. The ratio, $B_t / B_0$, is probably diverse among different galaxies as shown in \citep{2019Galax...8....4B}. For example, gas accretions and supernova explosions in galaxies induce more turbulence in numerical simulations \citep{2012MNRAS.422.2152B, 2013A&A...560A..87S}, and NGC 891 is known to have a specific star formation rate that is 2-3 times higher than in the Milky Way.  

In our model, we can add to a fraction of the gas a vertical magnetic field that could be caused by supernova explosions and associated blowouts. This effectively reduces the net fractional polarization by canceling polarized emission from dusty gas containing the expected planar field when averaged over or large beam. Without far higher spatial resolution than our HAWC+ beam, it is not possible to clearly distinguish regions with a net vertical field from the more normal regions with a planar field. We also found that reducing the intrinsic maximum polarization from 9\% to 5.5\% will make our model results (including turbulence) comparable to observations. The maximum polarization depends on the shape and composition of dust grains \citep{2015ARA&A..53..501A}, and dust polarization models from \citet{2018A&A...610A..16G} have maximum polarization from 5 to 10 \% at $154\micron$ depending on dust composition. However, a NIR study \citep{2000AJ....120.2920J} found that NGC891 has significantly lower fractional polarization compared to other nearly edge-on galaxies such as NGC 4565. Unless NGC891 has a very different dust composition from other galaxies, there is no clear reason to use a lower maximum polarization in our models. Note that \citet{2021ApJ...906..115T} show that the polarizing power of dust can depend on dust temperature, but they consider a much wider range in temperature (molecular cloud cores to highly radiated regions) than we sample with our FIR observations.

\section{Conclusion} \label{sec:conclusion}

Using our FIR observations in NGC 891 we find the following:

\begin{enumerate}

  \item The inferred magnetic field geometry near the galactic midplane is closely parallel to the plane of the disk and shows a decrease in fractional polarization near intensity peaks on either side of the nucleus. According to our models, the lower fractional polarization at these intensity peaks is likely due to the magnetic field lines being parallel to the LOS near the tangent of spiral features.
  
  \item The observations show significantly lower fractional polarization in the center of NGC 891 than expected for an edge-on spiral system. Our models rule out a simple spiral galaxy which has the magnetic field lines crossing our LOS through the nucleus, producing a maximum in fractional polarization, contrary to the observations. A model with the magnetic field aligned along the bar axis of a barred spiral, inclined from the plane of the sky, better fits the data. 

  \item Using the expected polarization efficiency of galactic dust, our models require a significant number of turbulence cells along a LOS in the magnetic field to reduce the model fractional polarization. In addition, the narrow dispersion in position angles in the plane of NGC 891 requires the model to have a significant number of turbulent regions along any LOS; otherwise, the model predicts too large a dispersion in position angles.
  
  \item To match the observed low fractional polarization, our model galaxy requires a stronger contribution of turbulent magnetic fields than inferred from observations of the Milky Way. The greater turbulence in the magnetic field in NGC 891 may be due to the higher active star formation rate compared to the Milky Way. Alternatively, a mixture of vertical and horizontal magnetic fields in the model disk in addition to the expected amount of turbulence will match the data. A much lower intrinsic polarization for the dust grains also works, but we consider this highly unlikely.

  \item The inferred magnetic field geometry in the disk is very closely aligned with the disk plane, yet there is one location where the magnetic field is clearly perpendicular to the galactic plane. This vertical polarization feature extends up to at least $\sim$2kpc into the halo in our FIR map. This location coincides with a NIR polarization null point and the location of dusty vertical cones extending off the disk seen in optical observations that are associated with regions of bright ionized emission. There has likely been enhanced star formation at this location on the near side of the disk causing a blowout that had dragged the magnetic field into a net vertical orientation. 
  
  \item  The inferred magnetic field geometry shows more complexity off the plane into the halo than in the disk. Compared with radio observations, our FIR observations show significant dispersion in the distribution of inferred magnetic field orientations, especially in the northeast and southwest regions outside the galactic plane. There is no clear signature of vertical fields off the plane and into the halo except at the NIR polarization null point, but the unexpectedly low fractional polarization in the halo is best explained as a mixture of vertical and horizontal magnetic fields that partially cancel in the net polarization.

\end{enumerate}

\begin{acknowledgments}
This research has made use of the NASA/IPAC Infrared Science Archive, which is funded by the National Aeronautics and Space Administration and operated by the California Institute of Technology. Based on observations made with the NASA/DLR Stratospheric Observatory for Infrared Astronomy (SOFIA). SOFIA is jointly operated by the Universities Space Research Association, Inc. (USRA), under NASA contract NNA17BF53C, and the Deutsches SOFIA Institut (DSI) under DLR contract 50 OK 2002 to the University of Stuttgart. Financial support for this work was provided by NASA through award AOR 09\_0067 issued by USRA.
\end{acknowledgments}


\vspace{5mm}
\facilities{IRSA, SOFIA (HAWC+)}


\software{APLpy \citep{aplpy2012, aplpy2019}, Astropy  \citep{2013A&A...558A..33A,2018AJ....156..123A}, Matplotlib \citep{Hunter:2007}, NumPy \citep{harris2020array}, SciPy \citep{2020SciPy-NMeth}, \texttt{emcee} \citep{2013PASP..125..306F}
          }



\appendix
\counterwithin{figure}{section}
\counterwithin{table}{section}

\section{Modelling synthetic images} \label{app:model}

Our goal is to create a simple model galaxy with a spatial distribution of warm dust that is similar to well observed face-on disk galaxies. The thermal radiation from this dust can then be integrated along a LOS corresponding to an edge-on view and compared to our surface brightness map of NGC 891 at $154~\mu m$. We will use mathematical descriptions of spiral and barred spiral galaxies that are similar to forms in the literature to maintain some connection with those studies and the parameters they use. These formulations have more parameters than we can reliably constrain with our data, but since we are primarily interested in modeling the magnetic field geometry that will thread through the density morphology, small variations in the density parameters will not matter.

\subsection{Density distribution of model galaxies} \label{app:density}

Modeling the density distribution of a spiral galaxy is commonly composed of three components: a central region (often a bulge), spiral arms or a bar plus spiral arms, and a diffuse disk in which the spiral structure is embedded. The vertical density profile of these components is usually modeled as exponential with a scale height (\textit{z}). The radial density is also an exponential or an exponential power-law profile as a function of distance from the center projected on the galactic plane (\textit{R}). Spiral arms can be defined with one pitch angle, assuming a logarithmic spiral structure.

We started with the equations for dust density distribution in \citet{2012ApJ...746...70S} and modified them to better suit our effort to model our observations at FIR wavelengths. The equations used to define the neutral and molecular hydrogen number density distributions, $\rho_{HI+H_2}$, in our models are described as follows:

\begin{align*}
\begin{split}
\rho_{HI+H_2, All}(R,\phi,z) &\propto  A_{diffuse}\times \exp(-(z/zd_{diff})) [\textrm{Diffuse disk}(R)] \\
&\qquad\qquad\qquad + \exp(-(z/zd_{thin}))[\textrm{Center}(R) + \textrm{Spiral~Structure}(R,\phi)]
\end{split} \\
\textrm{Diffuse disk}(R) &= \exp(-(R/R_d)) \\
\textrm{Center}(R) &= \exp(-(R/R_c)^{1/n_c}) \\
\textrm{Spiral Structure}(R,\phi) &= ws \times \exp(-(R/R_{s})^{10}) \times \textrm{Spirals}_{w/~or~w/o~bar}(R,\phi)\\
\textrm{Spirals}_{w/o~bar}(R,\phi) &= [\textrm{cos}(\ln{R}/\textrm{tan}(pitch) +PA - \phi)]^{20} \\
\textrm{Spirals}_{~w/~bar}(R,\phi) 
 &= wb \times [\textrm{cos}(PA - \phi)]^{20(R/R_{bar})}\qquad\qquad\qquad\qquad\qquad\qquad\textrm{where}~R \le R_{bar} \\
 &= [\textrm{cos}(\ln{R}/\textrm{tan}(pitch) - \ln{R_{bar}}/\textrm{tan}(pitch) + PA - \phi)]^{20}~\quad\textrm{where}~R > R_{bar} \\
\end{align*}

For the diffuse dusty disk, a single disk with an exponential profile both in radius and in height is commonly used \citep[e.g.,][]{2000A&A...353..117M, 2008A&A...490..461B, 2012ApJ...746...70S}. However, this formulation does not match our observations for NGC891, so we added a central component with a variable exponent in the term for the radial distribution. For simplicity, we assumed a nearly constant amplitude for the spiral arms within radius $R_s$, determined by amplitude $ws$. The multiplicative term $\exp(-(R/R_{s})^{10})$ forces the spiral arms to vanish quickly beyond $R_s$. The width of the spiral arm is determined by the exponent of a cosine function, similar to modeling by \citet{2012ApJ...746...70S}, and we fixed the exponent at 20 to ensure a narrow spiral arm and these arms are not resolved within our large beam.

For a barred spiral galaxy, the primary difference from a model spiral galaxy is the combined structure of a bar with spiral arms starting at the end of the bar instead of spiral arms alone. We assumed a straight bar and the spiral arms extended from a bar as logarithmic spirals. The variable $wb$ is used to define the difference in amplitude between the bar and the spiral arms. The exponent $20(R/R_{bar})$ within a bar region instead of just a value of 20 used for the pure spiral case ensures the bar will not become too thin as it approaches the central region.

The above equations have been normalized by the central component. So, we have to scale the values by deriving the dust emission for an edge-on view from the modeled neutral and molecular hydrogen gas column density, N(HI+H$_2$), and comparing it with our observation at FIR wavelengths. The conversion between FIR surface brightness and N(HI+H$_2$) is calculated with the modified blackbody function in the optically thin limit, $I_{\nu} = \tau_{\nu} B_{\nu}(T)$. $B_{\nu}(T)$ is the blackbody radiation, and the optical depth, $\tau_{\nu}$, is equivalent to $\mu m_H \textrm{N(HI+H}_2) \kappa_{\nu_0} (\nu / \nu_0)^{\beta}$ in the optically thin limit. The dust temperature, $T$, in NGC 891 is assumed as 23 K uniformly based on \citet{2014A&A...565A...4H}. If one assumes all the hydrogen associated with dust emission is molecular, appropriate in molecular clouds, $\mu$ will be 2.8 \citep[e.g.,][]{2013ApJ...767..126S}. Since the ratio of neutral to molecular hydrogen number density is diverse in galaxies, we use $\mu = 1.36$ \citep{1983QJRAS..24..267H}. For the opacity, $\kappa_{\nu_0}$, we assume 0.1 cm$^2$ g$^{-1}$ at $\nu_0$ = 1200 GHz \citep{1983QJRAS..24..267H}. The modeled dust emission is convolved with a Gaussian kernel with a scale of 13.6\arcsec (the HAWC+ $154\micron$ FWHM beam) and projected onto a synthetic observation grid. The model and our observations are compared in the center of the galaxy, and the modeled values are adjusted. Note that, with the assumption of constant temperature, N(HI+H$_2$) is directly proportional to $I_{\nu}$. We quantify the translation from dust emission to N(HI+H$_2$) in order to make comparisons with other work and quantify the modeling of turbulent cells in Appendix \ref{app:turbcells}.

To determine the parameters in the equations for the density distributions, we first applied the Markov chain Monte Carlo (MCMC) method using \texttt{emcee} \citep{2013PASP..125..306F}, searching for probable parameters to reproduce the observations. The searched variables are represented in Table \ref{table:density_param}. To reduce the variables, we fixed scale heights as 0.3 kpc and 0.1 kpc in the diffuse disk and the central and spiral components, respectively. For a barred galaxy, the position angle of the bar axis, $PA$, is set as -0.3 based on our polarimetry data, not the intensity image. This parameter was difficult to constrain in our MCMC method, given the limitation of an edge-on view. We ran 5000 steps with 200 chains and used the result from the last 500 steps to determine probable values for parameters. Due to so many parameters and the correlations between parameters, the MCMC chains converge very slowly. Although the results may not have converged properly yet, the 68\% confidence intervals of the parameters in the diffuse disk and a central part are less than 5\%. Note that the model fractional polarization in the center, which is shown in Figure \ref{fig:model}, depends on the contribution of the bar for the barred spiral model. We determined the parameters for the diffuse disk and central components from the results of the MCMC and then made small adjustments to the parameters for the barred spiral components to match the observed polarization in the center.

The final parameters are given in Table \ref{table:density_param}. Our model galaxies are designed to have the approaching side in the northeast, as shown in previous studies \citep[e.g.,][]{2007A&A...471L...1K}. Note that the northeast and the southwest portions of the disk are solved separately because of asymmetry in the location of the two bright regions to either side of the nucleus.

\begin{figure}
    \centering
    \includegraphics[width = \linewidth]{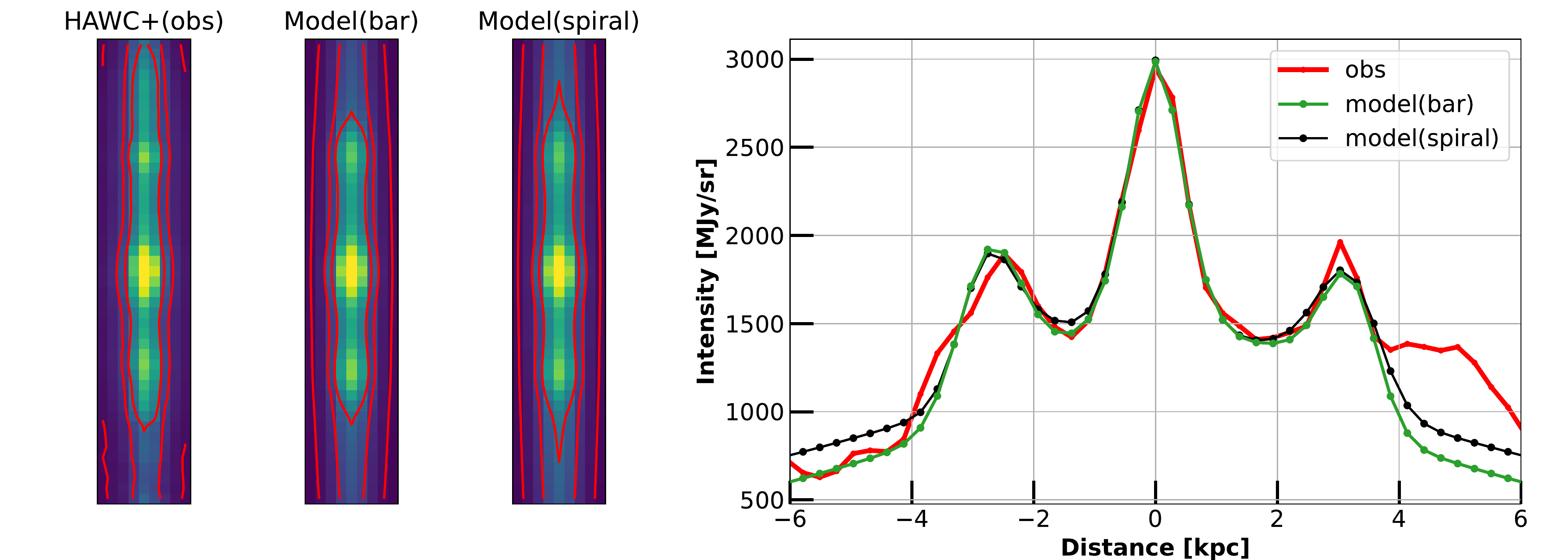}
    \caption{Color maps show the HAWC+ observation at $154\micron$ and the edge-on view of modeled galaxies. The pixel size and the beam FWHM are $6.8\arcsec$ and $13.6\arcsec$ for both the observation and the models. The red contours indicate 100, 500, and 1000 MJy/s. The total intensity profile along the midplane of the galaxy is plotted in the right panel.}
    \label{fig:density_model}
\end{figure}

\begin{figure}
    \centering
    \includegraphics[width=.9\linewidth]{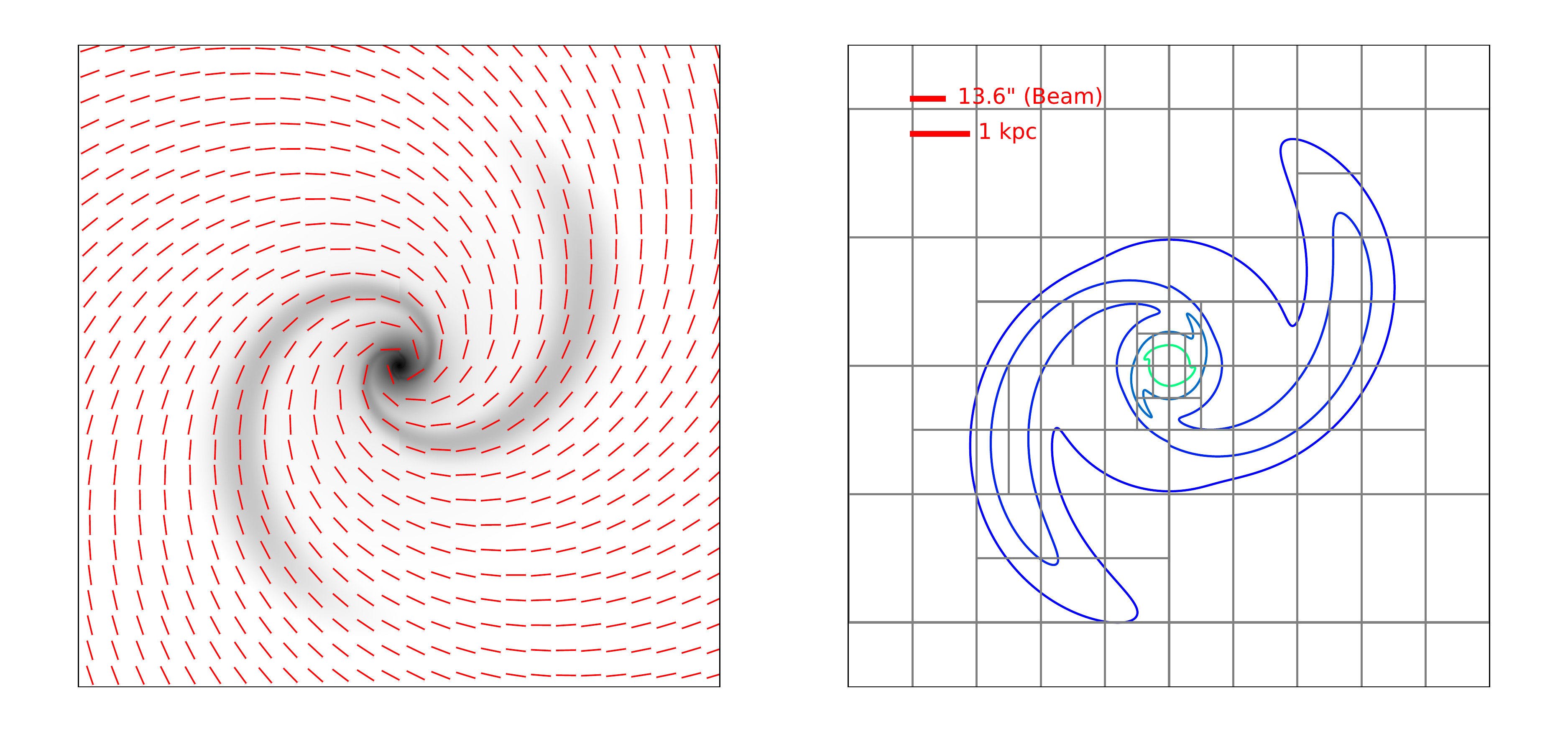}
    \includegraphics[width=.9\linewidth]{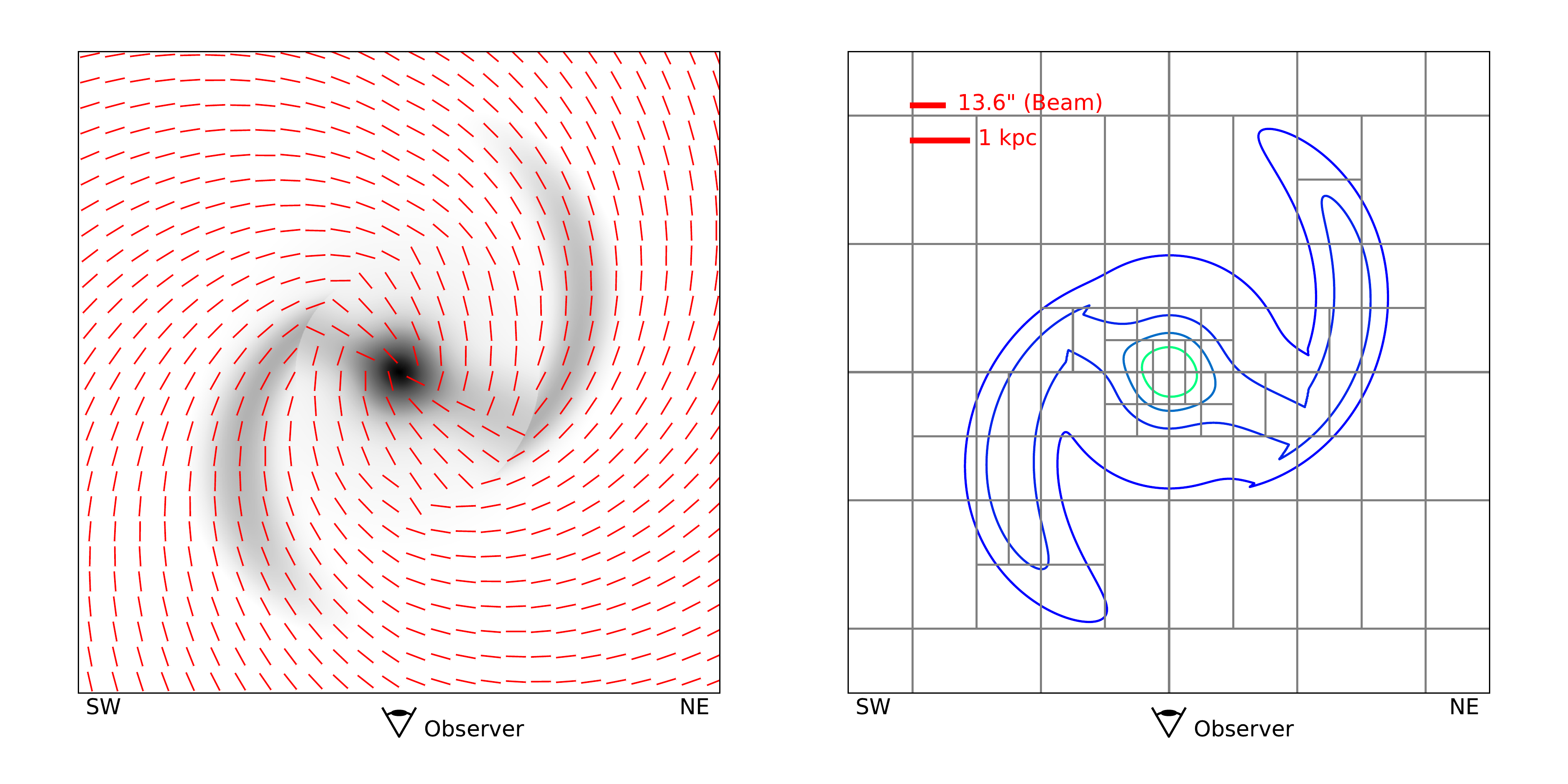}
    \caption{Visualizations of the ordered magnetic field geometry (left) and turbulence cells for turbulent magnetic fields in the galactic mid-plane (right). A spiral (barred spiral) galaxy is in the upper (bottom) panel. The size of a map is 11.75 kpc. The gas density distribution without beam convolution is shown as a gray map in a log scale on the left. Contours on the right represent hydrogen column density, N(HI+H$_2$), of 1.25, 2.5, 5.0, and 10.0 $\times 10^{21} \textrm{cm}^{-2}$. The turbulence cells shown here are when $\mathcal{N}_{th}$ is 10$^{64}$, the largest number among our trials (minimum number of turbulent cells), is used for clarity. The observer is located at the bottom of the images. The right (left) side is the northeast (southwest) of NGC891 projected on the plane of the sky.}
    \label{fig:model_param}
\end{figure}

Our model images with an edge-on view and the observed image at $154\micron$ are shown in  Figure \ref{fig:density_model}. The observed total intensity profile near the galactic midplane is well reproduced by both a spiral and a barred spiral model. However, the disk beyond 4kpc in the north is significantly brighter than the model and is inconsistent with either an exponential power-law or an exponential disk. The excess emission in the northern disk is also observed in Herschel 100 and $160\micron$ and Spitzer 24\micron \citep{2014A&A...565A...4H}. And this region is near where \citet{2014ApJ...786...41M} observed the Null polarization. Figure \ref{fig:model_param} depicts our model density distribution of the midplane viewed face-on.

Note that, because of many parameters for these components, the limit of the edge-on view, and low angular resolution at FIR wavelengths, it is hard to find a unique solution for a galaxy model. This point is as well shown by that our spiral model and barred spiral model give almost identical edge-on views. Also, our model assumes a constant temperature, but the variation of dust temperature within a dusty galaxy has been observed in face-on galaxies \citep[e.g.,][]{2012MNRAS.425..763G, 2012ApJ...755..165M}. However, we use constant temperature because we cannot determine the variation of temperature in an edge-on view and want to keep models simple.

\begin{table}
\centering
\begin{tabular}{@{} |c|c|cc|c|cc|c| @{}}
\hline
{}&\multicolumn{3}{c|}{Spiral galaxy}&\multicolumn{3}{c|}{Barred galaxy}&{unit}\\
\hline
{Diffuse Disk}&{$A_{diffuse}$}&\multicolumn{2}{c|}{0.015}&{$A_{diffuse}$}&\multicolumn{2}{c|}{0.023}&{--}\\
&{$R_d$}&\multicolumn{2}{c|}{7.27}&{$R_d$}&\multicolumn{2}{c|}{5.58}&{kpc}\\
&{$zd_{diff}$}&\multicolumn{2}{c|}{0.3}&{$zd_{diff}$}&\multicolumn{2}{c|}{0.3}&{kpc}\\
\hline
{Center}&{$R_c$}&\multicolumn{2}{c|}{0.24}&{$R_c$}&\multicolumn{2}{c|}{0.40}&{kpc}\\
{}&{$n_c$}&\multicolumn{2}{c|}{1.08}&{$n_c$}&\multicolumn{2}{c|}{0.80}&{--}\\
{}&{$zd_{thin}$}&\multicolumn{2}{c|}{0.1}&{$zd_{thin}$}&\multicolumn{2}{c|}{0.1}&{kpc}\\
\hline
{Spiral/Bar}&{$R_s$}&\multicolumn{2}{c|}{4.54}&{$R_s$}&\multicolumn{2}{c|}{4.62}&{kpc}\\
{}&{$zd_{thin}$}&\multicolumn{2}{c|}{0.1}&{$zd_{thin}$}&\multicolumn{2}{c|}{0.1}&{kpc}\\
{}&{$PA$}&\multicolumn{2}{c|}{1.21}&{$PA$}&\multicolumn{2}{c|}{-0.4}&{rad}\\
{}&{\textit{ws}}&\multicolumn{2}{c|}{0.07}&{\textit{ws}}&\multicolumn{2}{c|}{0.11}&{--}\\
{}&{--}&\multicolumn{2}{c|}{--}&{\textit{wb}}&\multicolumn{2}{c|}{0.57}&{--}\\
{}&{--}&(NE)&(SW)&{--}&(NE)&(SW)&{}\\
{}&{\textit{Pitch}}&{29.2}&{26.4}&{\textit{Pitch}}&{24.0}&{32.0}&{degr}\\
{}&{--}&{--}&{--}&{$R_{bar}$}&{2.57}&{1.90}&{kpc}\\
\hline
\end{tabular}
\caption{Parameters used to model the neutral and molecular hydrogen number density distributions. $PA$ is a position angle which spiral arms start from. To simplify the models, only a bar and spiral arms are defined individually in the northern and southern disks.} \label{table:density_param}
\end{table}

\subsection{Ordered magnetic fields} \label{app:orderedB}

The ordered component of the magnetic field in our model is assumed to be parallel to the galactic plane. The direction of the fields is not considered; only the orientation affects dust polarization. The ordered magnetic fields can be regular magnetic fields, anisotropic random magnetic fields, or both \citep{2013pss5.book..641B}. We assume the ordered magnetic field follows a spiral pattern characterized by a logarithmic pitch angle that also determines the hydrogen density distribution (see Appendix \ref{app:density}). A spiral structure in magnetic fields has been observed within many spiral galaxies \citep[e.g.,][]{2015A&ARv..24....4B, 2021ApJ...921..128B}. With a handful of spiral galaxies, \citet{2015ApJ...799...35V} found a clear correlation between the magnetic pitch angles and the pitch angles of spiral arms, and the difference between the two pitch angles is about 5\degr. \citet{2021ApJ...921..128B} observed smaller magnetic pitch angles in the inner radius and larger ones in the outer radius than morphological pitch angles in M51. To be simple, we assume that magnetic and morphological pitch angles are identical everywhere.

The magnetic field geometry for a barred spiral galaxy is defined separately in two parts. In the region within R$_{bar}$, we adopt the definition of magnetic field geometry used in Figure 18 of \citet{2005A&A...444..739B}. The authors observed two barred galaxies at radio wavelengths and found that magnetic fields are aligned with the leading edge of the bar in the outer bar region, as modeled in Figure 18. \citet{2021ApJ...923..150L} reported the comparison between radio and FIR polarization observations near the starburst ring in NGC1097, which is a barred galaxy with a starburst ring in the central region. The FIR observations show mostly constant magnetic field orientations in both the starburst ring and the outer bar, unlike the radio observations which show the twisted fields in the starburst ring. Thus, our model uses the magnetic field geometry inferred from the outer bar region up to the central part. Beyond R$_{bar}$, we make the magnetic fields oriented to have the same pitch angle with a spiral structure.
 
\subsection{Turbulence cells} \label{app:turbcells}

Current modeling methods for the magnetic fields in galaxies mostly focus on the Milky Way, and these works are reviewed in \citet{2016A&A...596A.103P} and \citet{2019Galax...7...52J}. Most studies generate isotropic random magnetic fields using a Gaussian random field with a Kolmogorov-like power spectrum. The power spectrum represents the magnitude of turbulent magnetic energy depending on the physical scale of the turbulence, and the random magnetic fields are generally generated in the grid of a constant cell size \citep[e.g.,][]{2010MNRAS.401.1013J, 2016A&A...596A.103P, 2016A&A...596A.105P}. Some models make an effort to match the models with observations by varying the amplitude of the random component depending on location in the Milky Way and adding an anisotropic random component \citep[see details in][]{2016A&A...596A.103P}.

Our study takes a somewhat different approach by varying the decorrelation size of the turbulence cells according to the gas density, and fixing the ratio of turbulent to ordered magnetic energy density to be constant everywhere. Our method focuses on examining how the spatial frequency (number of turbulent cells) of the turbulent component along a LOS and the amplitude of the turbulent component affect the integrated polarization in our observations. Having more turbulence cells within denser regions is commonly associated with a lower fractional polarization in higher molecular column density regions \citep{1992ApJ...389..602J, 2020A&A...641A..12P, 2022ApJ...936...92L}. 

Since the model contains a random component to the magnetic field in each turbulent cell, we run 500 realizations of the model and compute the median and dispersion in the fractional polarization and position angle for comparison with the observations. In the galactic plane of NGC 891, the data have high S/N and do not show a chaotic pattern either in fractional polarization or position angle. We will find that the very low fractional polarization will require a significant random component, but the low dispersion in fractional polarization and position angle will require an accompanying large number of turbulent cells for the model to match the low dispersion.

We define turbulence cells as consisting of several smaller model grid cells that  share the same random component. To produce more turbulence cells in denser regions, we require a turbulence cell to have approximately the same total number of hydrogen atoms and molecules, $\mathcal{N}(\textrm{HI+H}_2$), in the cell, as defined by the model parameter $\mathcal{N}_{th}$. Thus, the denser regions are divided into smaller size turbulent cells. \citet{1991A&A...248...23H} present that Faraday depolarization seen in NGC 891 can be explained by the turbulent magnetic field, of which the decorrelation scale changes with the thermal electron density to preserve the containing electron mass. In our model, the parameter $\mathcal{N}_{th}$ determines the number of turbulence cells in the model galaxy in the sense that decreasing $\mathcal{N}_{th}$ corresponds to a larger number of cells. 

Because of the strong drop-off in density along the vertical z-direction, turbulence cells in the thin disk component model, which is 1.18kpc thick, and the volume outside the thin disk (halo) are modeled separately. We implement a technique similar to an Adaptive mesh refinement in the thin disk. We first group the grid of 1536 × 1536 × 128 cells into 6 × 6 × 2 initial grid of turbulence cells, where $\mathcal{N}(\textrm{HI+H}_2$) is much greater than $\mathcal{N}_{th}$. The initial turbulence cells get divided into smaller cells until $\mathcal{N}(\textrm{HI+H}_2$) in the cell becomes close to the threshold value $\mathcal{N}_{th}$. If $\mathcal{N}(\textrm{HI+H}_2$) is above 6 $\times$ $\mathcal{N}_{th}$, the cell is volume-equally divided into 8 cells. For the cell above 3$\times~\mathcal{N}_{th}$ (1.5$\times~\mathcal{N}_{th}$), the cell is sliced into 4 cells (2 cells). Since the height of the starting turbulence cells is shorter than the width and length, the cells are divided only into 2 or 4 cells, not cutting parallel to the galactic plane, until the width, length, and height are the same. This volume-equally dividing process is repeated until all cells have $\mathcal{N}(\textrm{HI+H}_2$) less than 1.5$\times~\mathcal{N}_{th}$. As a result, most turbulent cells have a hydrogen number, $\mathcal{N}(\textrm{HI+H}_2$), between 0.5 and 1.5 $\times~\mathcal{N}_{th}$. During this process, a few turbulence cells with $\mathcal{N}(\textrm{HI+H}_2$) less than 0.5$\times~\mathcal{N}_{th}$ arise. These cells are merged with nearby cells. Outside the thin disk region where the density is low and does not vary much, the grid cells are volume-equally divided so that each turbulence cell will have $\mathcal{N}(\textrm{HI+H}_2$) close to $\mathcal{N}_{th}$.

This process creates an array of turbulent cells that each contain a roughly equal number of hydrogen atoms and molecules $\mathcal{N}(\textrm{HI+H}_2$). The right panels of Figure \ref{fig:model_param} illustrate how the derived turbulence cells are distributed in the galactic mid-plane in the case of $\mathcal{N}_{th}=10^{64}$. We see that the regions with higher volume density have physically smaller turbulence cells and that the LOS toward the galactic center on an edge-on view passes through more turbulence cells. The value for $\mathcal{N}_{th}$ used in Figure \ref{fig:model_param} is larger than we will use in the modeling to reduce the number of turbulent cells in the plot so that the reader can see the effect of our technique.

In all results in this paper but the case with $B_t / B_0 = 1.5$, we use parameter $\mathcal{N}_{th}=10^{63}$. This value is equivalent to a gas mass of 1.1$\times 10^6 M_\odot$ per turbulent cell. This makes roughly 10 times the number of turbulence cells in a galaxy than the case shown in Figure \ref{fig:model_param}. In our barred spiral model with $\mathcal{N}_{th}$ of $10^{63}$, the mean column density for each turbulent cell along the LOS through the galactic center is N(HI+H$_2$) = $2.2 \times 10^{21} \textrm{cm}^{-2}$. In terms of visual extinction, $A_{V} \sim 1$ is approximately equivalent to N(HI+H$_2$) = $2\times10^{21} \textrm{cm}^{-2}$ \citep{1978ApJ...224..132B, 2003ApJ...598.1017D}. This value is comparable with the result from \citet{1992ApJ...389..602J}, who found that interstellar polarization in the Milky Way could be explained by a model having a series of turbulent cells with extinction $A_{V}\sim 1$ in each, and a ratio of turbulent to the ordered field strength of B$_t$/B$_0$ = 1.

\begin{figure}
    \centering
    \includegraphics[width = .9 \linewidth]{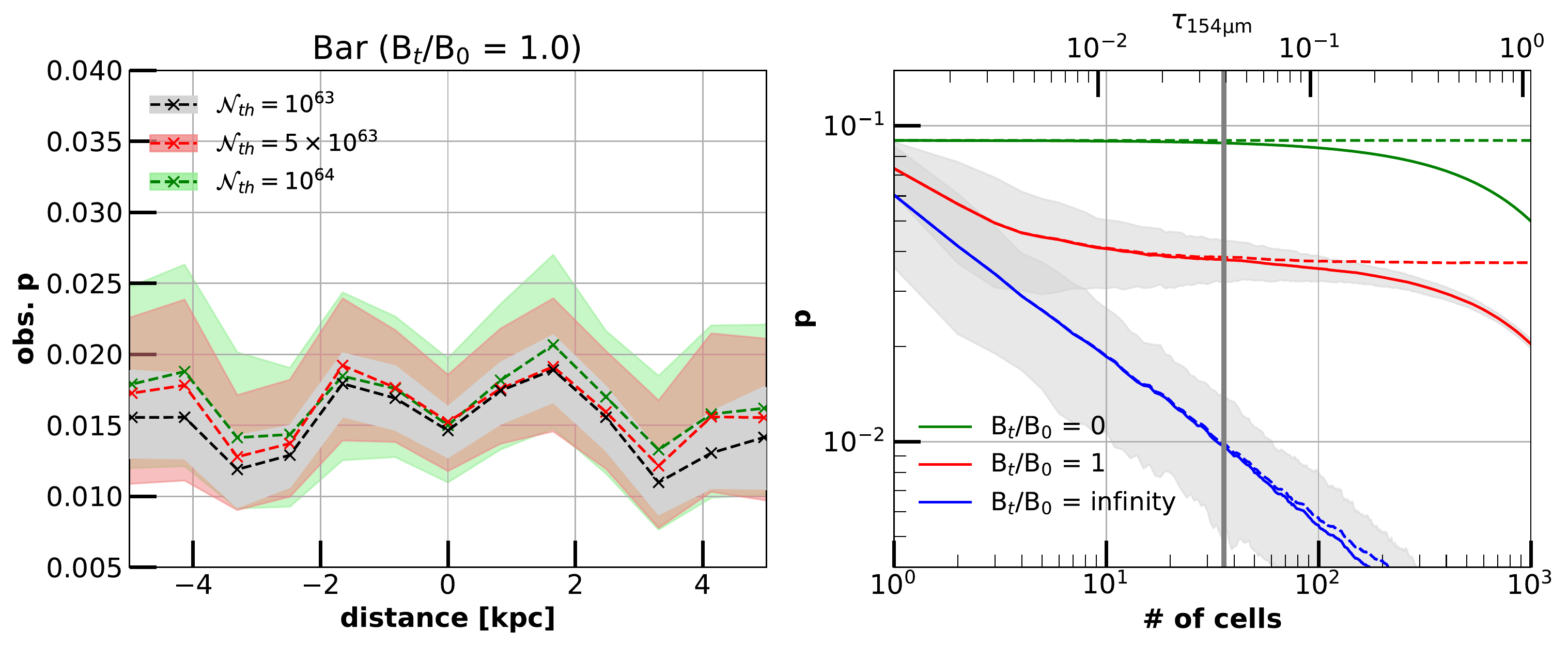}
    \caption{\textbf{Left}: Same as the barred galaxy in the right panel of Figure \ref{fig:model}, but with various threshold numbers ($\mathcal{N}_{th}$) for turbulence cells. Different colors show different $\mathcal{N}_{th}$. The numbers of turbulence cells along a LOS through the center are 32, 15, and 10, respectively from the smallest to the largest $\mathcal{N}_{th}$. The shaded regions indicate the model dispersion in fractional polarization. \textbf{Right}: Change of fractional polarization while passing through turbulence cells. Hydrogen column density, N(HI+H$_2$), of each cell is $2.0\times10^{21} \textrm{cm}^{-2}$. Solid lines represent the fractional polarization due to the combination of dichroic extinction and emission, calculated with the equations of transfer in \citet{2014ApJ...797...74D}, but dashed lines ignore the term for dichroic extinction in the equations. The grey shade shows the spread of 60\% of 500 simulations. The vertical line is at an optical depth of 0.039, which occurs at the center of our modeled galaxy.}
    \label{fig:turb_model}
\end{figure}

To investigate the effect of using different numbers of turbulence cells, we examined our barred spiral model with $\mathcal{N}_{th}$ of 1.0$\times 10^{63}$, 5.0$\times 10^{63}$ and 1.0$\times 10^{64}$. The model results are shown in the left of Figure \ref{fig:turb_model}. The figure corresponds to the barred galaxy model in the right panel of Figure \ref{fig:model}, but using several different values for $\mathcal{N}_{th}$. We can see that adding more turbulence cells does not significantly affect the median fractional polarization from 500 simulations. However, the spread in fractional polarization from the 500 realizations does decrease with an increase in the number of turbulence cells. This shows that while the fractional polarization along a LOS will saturate as the number of turbulent cells is increased, the dispersion among the simulations will continue to decrease.

In most of our results in this paper, we compare the observations with the model having $\mathcal{N}_{th}$ of 1.0$\times 10^{63}$ (the finest turbulence cells). Our observation results shown in Figure \ref{fig:model} and \ref{fig:model_3}, which have high S/N and closely follow the median values of our simulations, imply the significant number of turbulence cells are necessary for the model. Note that we use $\mathcal{N}_{th}=5\times10^{62}$ in the case with $B_t / B_0 = 1.5$. The stronger turbulence makes the larger dispersion in fractional polarization and position angles, so we need to compensate for this effect with more turbulence cells.

To better illustrate the effect of increasing the number of turbulent cells, we ran a simple simulation with one LOS through cells, each with A$_V$ = 1, and varied the total number of cells. We used three values for B$_t$/B$_0$, 0, 1, and \textit{infinity}. In each case, we run 500 times realizations, and the median values (solid) and the spread (shade) of the results are seen in the right hand panel in Figure \ref{fig:turb_model}. The optical depth at $154\micron$ corresponding to the number of turbulence cells is labeled on the upper axis. For the case with no ordered component, only the random component, the fractional polarization continues to drop with optical depth as roughly $\tau^{-0.5}$ \citep{2015AJ....149...31J}. For the case with no random component, the fractional polarization remains constant until optical depth effects take over. For the intermediate case, the fractional polarization initially drops with optical depth due to the random component, then levels off (saturates) until large optical depths are again encountered. 

We can see from this result that for over a factor of 10 or so spread in optical depth, our model predicts a nearly constant fractional polarization, but decreasing dispersion in that value. Along such a LOS, the turbulent component of the magnetic field is partially averaged out, and the geometry of the ordered component is the primary contributor to the variations through the edge-on disk in our model. The vertical line in Figure \ref{fig:turb_model} corresponds to the optical depth in our barred galaxy model as viewed edge-on through the center. So, in the regime we are observing, the polarization is affected not by optical depth (in the optically thin regime) itself but by the number of turbulence cells. We should note that the observed optical depth at the center in our HAWC+ beam is $\sim$0.015, but the model optical depth through the area corresponding to the smallest turbulence cell (which is smaller than our observation beam size) is 0.039. Gridding and beam convolution dilute the maximum N(HI+H$_2$) from our model, hence the lower observed value.


\bibliography{biblist}{}
\bibliographystyle{aasjournal}



\end{document}